\documentclass[final,3p,times]{elsarticle}

\usepackage{amssymb}
\usepackage{amsmath}
\usepackage{tabularx}
\usepackage{xcolor,colortbl}
\usepackage{diagbox}
\usepackage{caption}
\usepackage{subcaption}
\usepackage{makecell}
\usepackage{placeins}

\begin{document}

\begin{frontmatter}\

\title{Floating Car Observers in Intelligent Transportation Systems: Detection Modeling and Temporal Insights}

\author[inst1]{Jeremias Gerner}

\affiliation[inst1]{organization={Technische Hochschule Ingolstadt},
            department={AI Motion Bavaria},
            city={Ingolstadt},
            country={Germany}}
\author[inst2]{Klaus Bogenberger}
\author[inst1]{Stefanie Schmidtner}

\affiliation[inst2]{organization={Technical University Munich},
            department={Chair of Traffic Engineering and Control},
            city={Munich},
            country={Germany}}

\begin{abstract}
Floating Car Observers (FCOs) extend traditional Floating Car Data (FCD) by integrating onboard sensors to detect and localize other traffic participants, providing richer and more detailed traffic data. In this work, we explore various modeling approaches for FCO detections within microscopic traffic simulations to evaluate their potential for Intelligent Transportation System (ITS) applications. These approaches range from 2D raytracing to high-fidelity co-simulations that emulate real-world sensors and integrate 3D object detection algorithms to closely replicate FCO detections. Additionally, we introduce a neural network-based emulation technique that effectively approximates the results of high-fidelity co-simulations. This approach captures the unique characteristics of FCO detections while offering a fast and scalable solution for modeling. Using this emulation method, we investigate the impact of FCO data in a digital twin of a traffic network modeled in SUMO. Results demonstrate that even at a 20\% penetration rate, FCOs using LiDAR-based detections can identify 65\% of vehicles across various intersections and traffic demand scenarios. Further potential emerges when temporal insights are integrated, enabling the recovery of previously detected but currently unseen vehicles. By employing data-driven methods, we recover over 80\% of these vehicles with minimal positional deviations. These findings underscore the potential of FCOs for ITS, particularly in enhancing traffic state estimation and monitoring under varying penetration rates and traffic conditions.
\end{abstract}

\begin{keyword}
Floating Car Observers \sep Intelligent Transportation Systems \sep Traffic State Estimation \sep Extended Floating Car Data
\end{keyword}

\end{frontmatter}
\section{Introduction} \label{sec:Introduction}
Perceiving the current traffic state through diverse data sources is fundamental to developing intelligent transportation systems (ITS). By capturing, processing, and analyzing real-time traffic data, ITS enhances transportation networks' efficiency, safety, and sustainability. Key examples include adaptive traffic signal control, which dynamically adjusts signal timings based on real-time conditions \cite{jing2017adaptive}; eco-driving assistance systems that provide drivers with feedback to optimize fuel efficiency and safety \cite{huang2018eco, schlamp2023glosa}; and intelligent routing systems that leverage traffic data to determine optimal routes aligned with specific objectives \cite{tyagi2022routing}. Among the various traffic data sources, induction loops remain the most widely available and utilized. Embedded in road surfaces, they provide only sparse data, as they are installed at dedicated locations, often limited to intersections or key road segments. This spatial restriction means they cannot offer continuous network-wide coverage, leaving large areas unmonitored. Furthermore, their data is macroscopic, capturing aggregated information such as vehicle counts or flow rates, but lacking detailed insights into individual traffic participants. Additional stationary sensors that provide more detailed information about individual road users can help here. So-called roadside infrastructure sensors include various sensors such as RGB cameras,  LiDAR, and RADAR, which can detect and localize other road users. This means that much more detailed information about individual road users can be obtained and made available to the ITS systems. The main problem with such sensors is the costs involved in installing and operating them on a large scale. In a 15-year life cycle cost analysis, \cite{2023kloekereconomic} estimates that for the city of Cologne, Germany, with a population of approximately one million, the cost of complete coverage with intelligent roadside sensors would exceed 1.3 billion Euros. This coverage includes a total of 8,610 nodes across urban, rural, and highway domains. Through Vehicle-to-Everything (V2X) communication, a new frontier in traffic data collection has emerged. V2X enables vehicles equipped with communication devices to exchange information with other vehicles and entities, such as infrastructure or centralized systems. When integrated with technologies like GPS and other localization systems that achieve centimeter-level accuracy \cite{2023shanlocalization}, vehicles can transmit their current position as Floating Car Data (FCD). Additionally, supplementary information such as speed and acceleration can also be shared. Unlike stationary sensors, FCD provides a dynamic and continuous data stream that spans the entire road network, offering more detailed insights into traffic conditions and flow patterns. Despite its advantages, FCD faces challenges. A primary limitation is the penetration rate ($p$), which is defined as the proportion of vehicles within the traffic network that are capable of and willing to share their data. Low penetration rates can result in sparse or unrepresentative data made available. Moreover, FCD predominantly captures data about vehicles, leaving vulnerable road users (VRUs)—such as pedestrians and cyclists—largely unaccounted.
\newline
\newline
Many vehicles equipped with V2X communication and localization technologies are also fitted with sensors capable of perceiving their surroundings. Similar to roadside sensor systems, these include LiDAR, RADAR, and RGB camera systems that are permanently installed in the vehicles. These sensors are essential for enabling Advanced Driver Assistance System (ADAS) functionalities. The level of automation is classified into different stages, as defined by \cite{SAEJ3016}, ranging from Level 0 (no automation) to Level 5 (full automation). Depending on the ADAS level of the vehicle, sensors are installed to meet the corresponding requirements. Starting from Level 3 (conditional automation), vehicles are equipped with sensors that can detect and localize other traffic participants, such as pedestrians, cyclists, and other vehicles. This capability allows each vehicle to gather additional information about its environment and other road users. Moreover, research in Connected Automated Vehicle (CAV) functionalities explores how this data can be shared between vehicles to enhance ADAS capabilities \cite{elliott2019cav}. On a network-wide scale, there is potential to utilize the additional information gathered by each vehicle's sensors. Similar to the concept of FCD, V2X communication enables not only the sharing of a vehicle's individual data but also the transmission of information collected through its sensors, specifically the detected and localized traffic participants. This approach introduces the concept of Extended Floating Car Data (xFCD). Vehicles that transmit not only their individual data but also xFCD are referred to as Floating Car Observers (FCO). This extension can enhance the quality and granularity of traffic data available for ITS systems.
\newline
\newline
Recent works in the area of FCO have demonstrated their potential for traffic state estimation and analysis. For instance, \cite{ma2021high} utilized the detection capabilities of connected vehicles to estimate key traffic state variables such as flow, density, and speed. Similarly, \cite{florin2022real} proposed a method to estimate highway traffic density using FCO data, incorporating traffic observations from both travel directions. Furthermore, the estimation of macroscopic fundamental diagrams using FCOs was introduced in \cite{zhang2023novel}. These studies predominantly focus on processing FCO information for macroscopic traffic analysis. However, they often overlook the need for accurate modeling of detections and the characteristics of real-world systems. In this work, we aim to address this gap by exploring various modeling techniques for FCO detections within a microscopic traffic simulation environment. Beyond detection modeling, we investigate how previously detected traffic participants can contribute to the traffic state information, even if they are currently not detectable by one of the FCOs anymore. While the proposed evaluation processes in this work are applicable to a diverse set of traffic participants, we focus on a vehicle-only setting to concentrate on the methodology itself.

\section{Vehicle Detection and Localization}\label{sec:background_detection}
The key advantage of FCO over traditional FCD lies in its ability to detect and localize other traffic participants using advanced onboard sensors, thereby providing richer information for traffic monitoring and control. Those sensors include RGB cameras, and LiDAR. RGB camera sensors capture high-resolution images of the environment, offering rich visual information that supports semantic object detection. This includes recognizing object classes, such as vehicles or pedestrians, and extracting detailed contextual information about the scene. RGB cameras, therefore, are essential for detecting and identifying objects with high semantic granularity. LiDAR sensors complement this by generating precise 3D point clouds of the surrounding area. While LiDAR data lacks inherent semantic information, its primary strength lies in providing highly accurate depth measurements, allowing for reliable spatial localization and geometric reconstruction of detected objects. This depth information is critical for distinguishing between objects and accurately estimating their positions in three-dimensional space. In addition, RADAR sensors may enhance 3D object detection by contributing robust spatial measurements under conditions where RGB cameras and LiDAR face challenges, such as in low-light environments or adverse weather conditions. From the raw sensor data alone, traffic participants cannot be directly detected or localized. This capability emerges only when models are applied to process the raw data into structured, interpretable outputs. In contemporary systems, these models are predominantly based on deep learning, leveraging advanced architectures such as Convolutional Neural Networks (CNNs) to extract features and predict the locations and identities of traffic participants. Specifically, in the context of 3D object detection, this process can be formalized as:

\begin{equation}
\mathcal{B} = f(\mathcal{S}),
\end{equation}

where $\mathcal{B}$ represents a set of $N$ detected 3D traffic participants in the scene, $f$ denotes the 3D object detection model, and $\mathcal{S}$ refers to the raw sensor data utilized by the model. For these models to achieve high performance, they must first be trained on large, diverse datasets that provide accurate ground truth labels $\hat{\mathcal{B}}$. Popular datasets for 3D object detection include the KITTI Dataset \cite{KITTI}, Waymo Open Dataset \cite{sun2020scalability}, and nuScenes Dataset \cite{nuscenes}. Among these, the KITTI Dataset is one of the earliest and most widely adopted benchmarks for 3D object detection research. It provides over 15,000 annotated bounding boxes, distinguishing between key classes such as cars, pedestrians, and cyclists. The dataset is generated using a stereo RGB camera system and a LiDAR sensor. A feature of the KITTI Dataset is its inclusion of difficulty levels for annotated objects, enabling more granular evaluation of detection models. These difficulty levels are determined by thresholds based on three criteria: the object's size in the image, the degree of occlusion, and the extent of truncation. Utilizing the KITTI dataset, different state-of-the-art 3D detection algorithms were trained and benchmarked on the dataset using either an RGB camera or a LiDAR. Additionally, multimodal architectures that fuse both RGB camera and LiDAR data have emerged, aiming to harness the complementary strengths of these sensors. These fusion-based models combine the semantic richness of RGB camera data with the spatial accuracy of LiDAR. The main metric used for evaluating the performance of 3D detection algorithms is the Average precision (AP). Either evaluated on the 3D bounding box or the Bird’s Eye View (BEV) representation of the predicted and ground truth bounding boxes (referred to as BEV AP), the AP measures detection performance by calculating the area under the precision-recall curve. For vehicles, a detection is considered valid if the Intersection over Union (IOU) between the predicted and ground truth bounding boxes exceeds a threshold of 0.7 \cite{KITTI} which is also referred to as AP@70.
\newline
\newline
Traditionally, the detections $\mathcal{B}$ obtained from 3D object detection algorithms are utilized locally within the perception-planning-action pipeline of modular autonomous driving systems. With advancements in V2X communication technologies, it has become feasible to regularly and reliably exchange larger data packets among different participants in the traffic system. In the context of Autonomous Driving (AD), this extends autonomous vehicles to Connected Autonomous Vehicles (CAV), which can enhance the operational safety of the vehicles. Consequently, vehicles equipped with ADAS functionalities likely incorporate corresponding V2X systems that allow the sharing of such information. Hence FCO systems do not incorporate additional cost in this regard. The information that is shared can be at different levels: raw sensor data, intermediate features from deep learning models, or lists of detected objects. In the context of FCOs in this work, we focus on the latter case—sharing lists of detected objects, reducing the need for additional processing. This approach involves disseminating the information gathered by different observers, together with their current positions obtained via different localization methods, to a central entity via RSUs within the traffic system. This central entity aggregates data about detected vehicles to generate microscopic traffic state information at specific points in time for the traffic network or relevant parts of it, which can then be utilized by different ITS algorithms.

\section{Modeling Floating Car Observers in Microscopic Traffic Simulations}\label{sec:detection_modeling}
\begin{figure}[h]
    \centering
    \begin{subfigure}{0.32\textwidth}
        \centering
        \includegraphics[width=\linewidth]{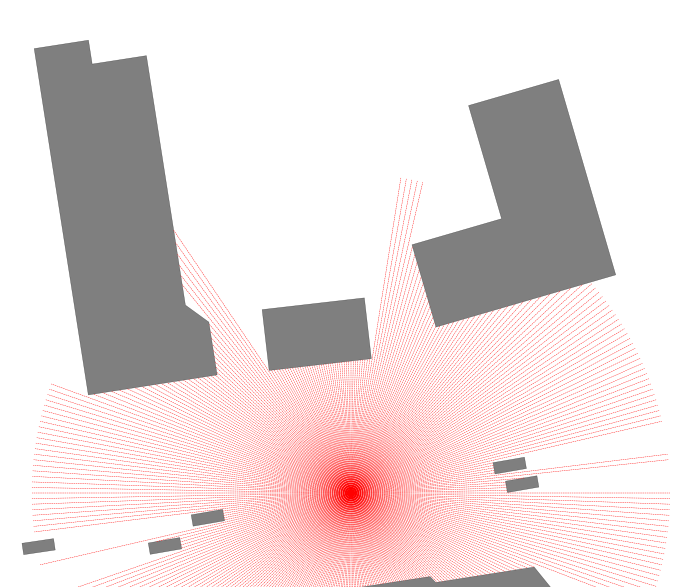}
        \caption{Visualization of the 2D raytracing detection modeling}
        \label{fig:2d_raytracing}
    \end{subfigure}
    \begin{subfigure}{0.32\textwidth}
        \centering
        \includegraphics[width=\linewidth]{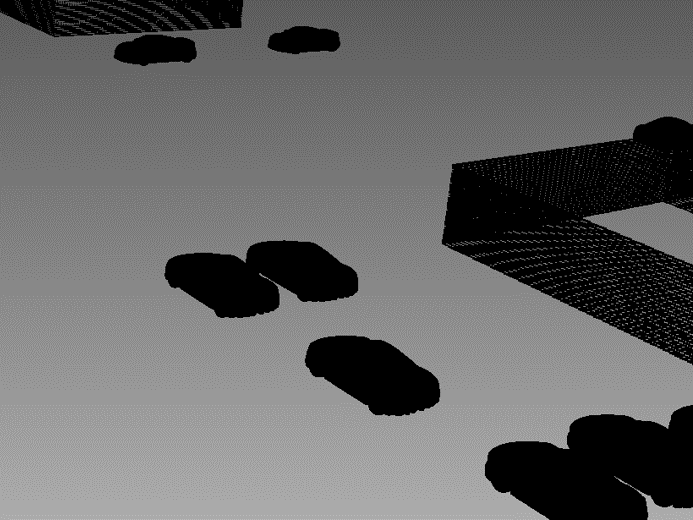}
        \caption{Visualization of the 3D point cloud used in the 3D raytracing detection modeling}
        \label{fig:3d-pointcloud}
    \end{subfigure}
    \begin{subfigure}{0.32\textwidth}
        \centering
        \includegraphics[width=\linewidth]{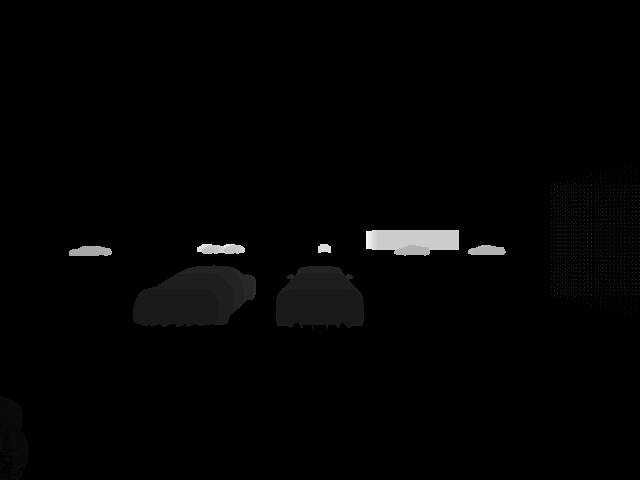}
        \caption{Depth image generated during the 3D raytracing detection modeling.}
        \label{fig:3d-depth}
    \end{subfigure}
    \caption{Visualizations of the 2D and 3D raytracing modeling approaches}
    \label{fig:main}
\end{figure}
The previous sections introduced the concept of FCOs, highlighting their potential for traffic analysis and their integration into traffic optimization algorithms. However, their real-world implementation remains largely unrealized. One potential reason for this gap is the need for a thorough evaluation of the benefits of the xFCD delivered by FCOs compared to FCD and traditional stationary sensors, such as induction loops. Implementing large-scale FCOs in real-world scenarios entails substantial costs and challenges, including concerns about maintaining minimum performance standards in traffic control, making direct examination of the concept in real-world studies less appealing. As an alternative, microscopic traffic simulations provide a promising and cost-effective approach for evaluating the concept in controlled environments. Microscopic traffic simulations such as SUMO \cite{SUMO2018} are detailed, agent-based models that replicate the behavior of individual vehicles and their interactions within a traffic system. These simulations are widely used in traffic research and control as they enable researchers and practitioners to study complex traffic dynamics under controlled conditions, making it possible to test and evaluate the impact of various traffic control strategies and technologies. Hence, those simulations can also be used to evaluate the potential benefits of using FCO information for various traffic analysis and traffic optimization algorithms. The basis for this evaluation is the need for an accurate estimation, within the microscopic simulation, of which vehicles can be detected by an FCO using its onboard sensors and 3D detection algorithms, as well as their precise positioning in 3D space. However, microscopic simulations are not inherently designed for such detailed estimations; they are typically highly abstract. For instance, standard microscopic traffic simulation frameworks do not allow for the direct attachment of sensors to individual vehicles, nor do they simulate the detailed physics or environmental factors required for realistic sensor performance modeling. It is, therefore, necessary to develop and apply detection models, which estimate the FCO detectability within the microscopic traffic simulation. I.e. given a single FCO in the traffic scene within a microscopic traffic simulation, we want to examine which vehicles would be detectable in a comparable real-world scenario. Scaling such detection modeling to all FCOs present within the traffic network enables to determine the effectiveness of the FCO approach. Formally, when we denote $V_t$ as the set of all vehicles in the traffic network at time $t$. Let $O_t \subseteq V_t$ be the subset of vehicles acting as FCOs. Each FCO can detect other vehicles in the network. The goal of detection modeling is to determine $D_t$, the set of detections available through the FCO system at time $t$, where each detection corresponds to a vehicle in $V_t$. 

\subsection{Two-Dimensional Raytracing}\label{subsec:2D_raytracing}
A straightforward method to approximate the detectability is a two-dimensional (2D) raytracing approach. This method utilizes information about traffic participants along with polygon data from environmental objects to create a BEV-like representation around the currently investigated FCO. In this approach, a predefined number of rays are emitted from the center of the observer vehicle, extending outward to a specified maximum range. Each ray checks whether and which object it intersects first, effectively determining the Field of View (FoV) of the current FCO. This method provides insight into the portions of the urban scenario that are occluded or visible to the observer, as demonstrated in \cite{Ilic2024raytracing}. A visualization of this process is shown in Figure \ref{fig:2d_raytracing}. Building on this foundation, we extend the implementation by incorporating a mechanism to count the number of rays that intersect with other traffic participants. Using this information, we can apply a threshold to classify vehicles as either detectable or undetectable which effectively implies occlusion or out of range. Thus, the 2D raytracing provides the first modeling technique for FCO detectability.
\newline

\subsection{Three-Dimensional Raytracing}\label{subsec:3D_raytracing}
In \cite{gerner2023sumodetector}, we introduced a more advanced method for determining which vehicles are detectable in a given traffic scenario based on the sensor setup, specifically focusing on RGB camera configurations. In this work, we refer to this method as three-dimensional (3D) raytracing. Within the 3D raytracing method, we first generate a three-dimensional point cloud representing the current traffic scene around the investigated FCO to evaluate detectability. This point cloud includes 3D representations of traffic participants and environmental objects. In our initial work, vehicles were represented as simple box-like structures; however, as shown in Figure \ref{fig:3d-depth}, more detailed and realistic 3D models can also be integrated into this process. Within this virtual 3D environment, various RGB cameras can be placed at varying positions and orientations on the FCO, replicating real-world RGB camera setups. These RGB cameras can be configured with specific settings, such as focal length and image resolution, to match hardware characteristics. Using Computer Vision (CV) techniques, the 3D points are projected onto the 2D image planes of these RGB cameras. This simulation replicates how RGB cameras perceive the environment, accounting for critical factors like occlusion, distance, and object size within the image. The process generates depth images with known object presence for each pixel within the depth image. An example depth image is shown in Figure \ref{fig:3d-depth}. By applying the well-established thresholds regarding object presence, occlusion and truncation from the KITTI dataset, introduced in Section \ref{sec:background_detection}, which is also used to train state-of-the-art 3D detection algorithms we can distinguish which vehicles within the depth image are detectable under the given traffic scenario.\footnote{In our previous work \cite{gerner2023sumodetector}, detections were only determined by the threshold of object presence, we extended this to all three characteristics of objected presence, occlusion, and truncation to fully cover the notion introduced in the KITTI dataset.} 

\subsection{Co-Simulation}\label{subsec:co-simulation}
\begin{figure}[tb]
    \centering
    \begin{subfigure}{0.42\textwidth}
        \centering
        \includegraphics[width=\linewidth, trim=0 75 0 0, clip]{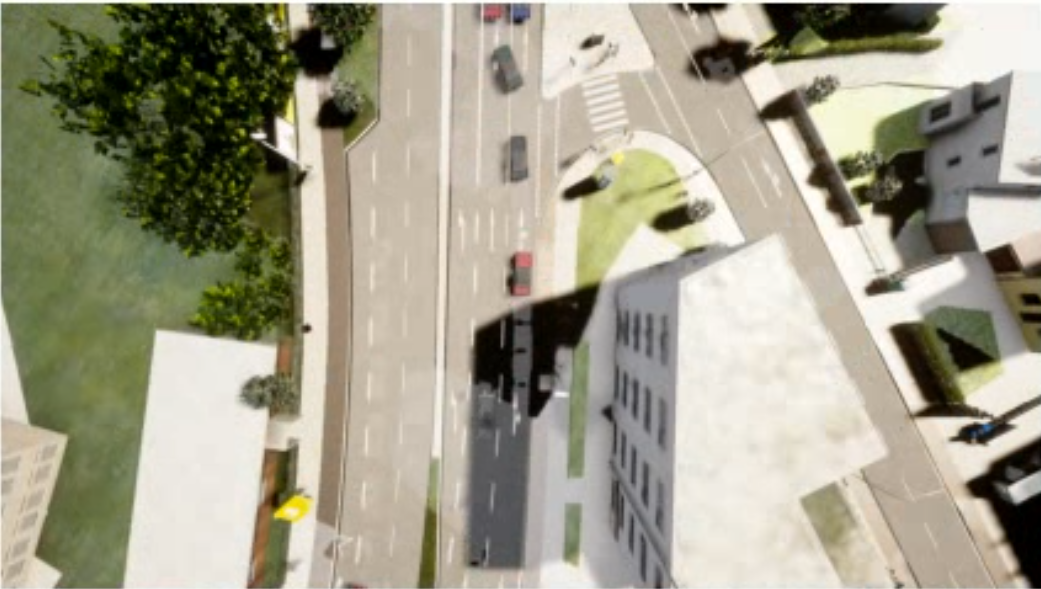}
        \caption{High-fidelity CARLA simulation of urban traffic in Ingolstadt, Germany.}
        \label{fig:carla_image}
    \end{subfigure}
    \hspace{0.03\textwidth}
    \begin{subfigure}{0.42\textwidth}
        \centering
        \includegraphics[width=\linewidth, trim=0 135 0 0, clip]{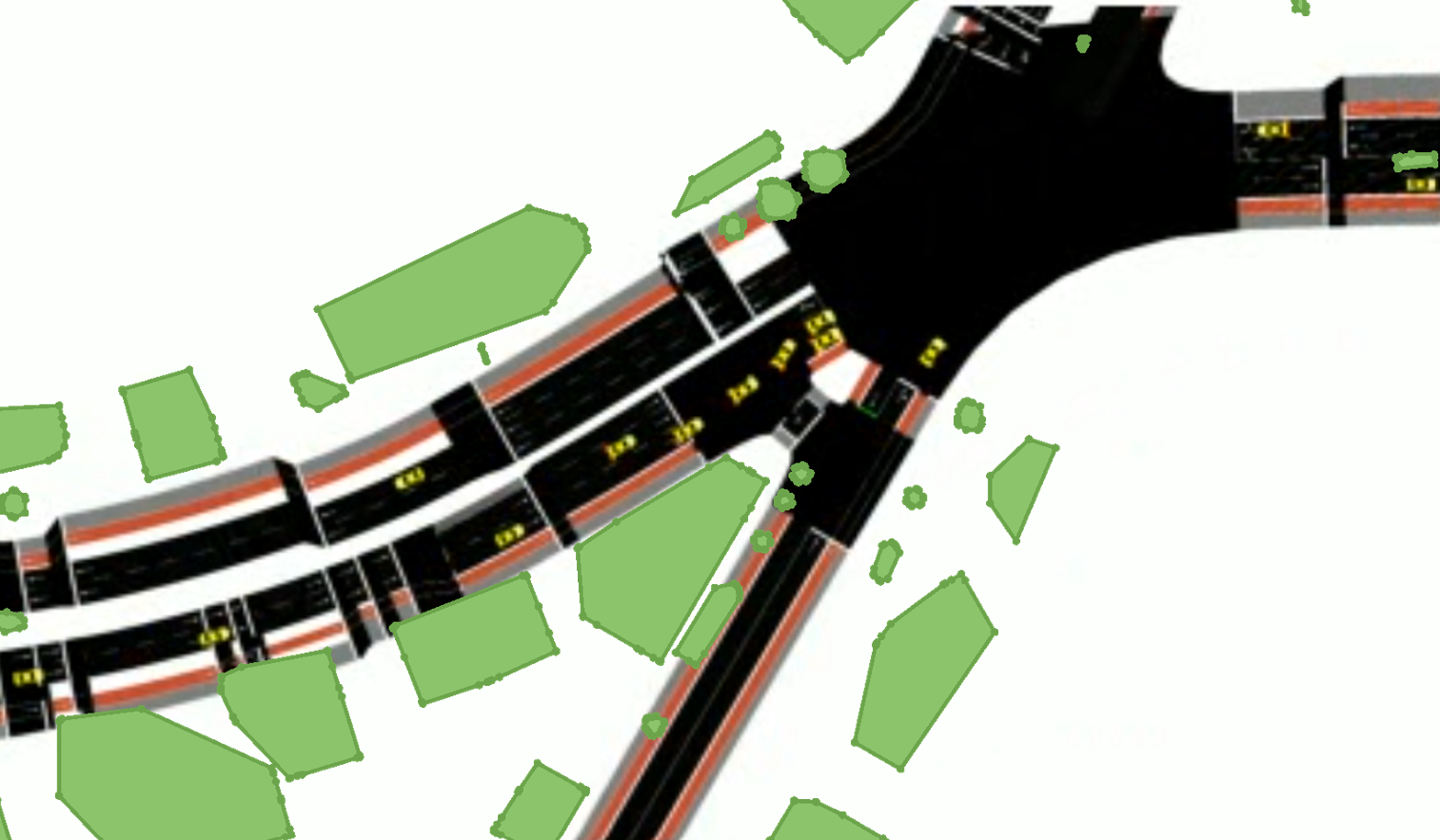}
        \caption{SUMO simulation of the identical urban traffic scene depicted in CARLA map (a), highlighting its low-level abstraction. Extracted polygons are shown in green.}
        \label{fig:sumo_image}
    \end{subfigure}
    \begin{subfigure}{0.42\textwidth}
        \centering
        \includegraphics[width=\linewidth]{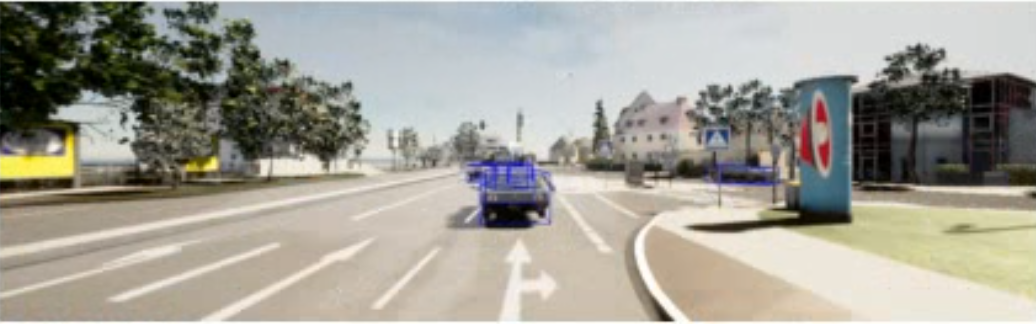}
        \caption{View of the front camera mounted onto a vehicle. The image is further processed with the ground truth 3D bounding boxes.}
        \label{fig:camera_image}
    \end{subfigure}
    \hspace{0.03\textwidth}
    \begin{subfigure}{0.42\textwidth}
        \centering
        \includegraphics[width=\linewidth, trim=0 119 0 118, clip]{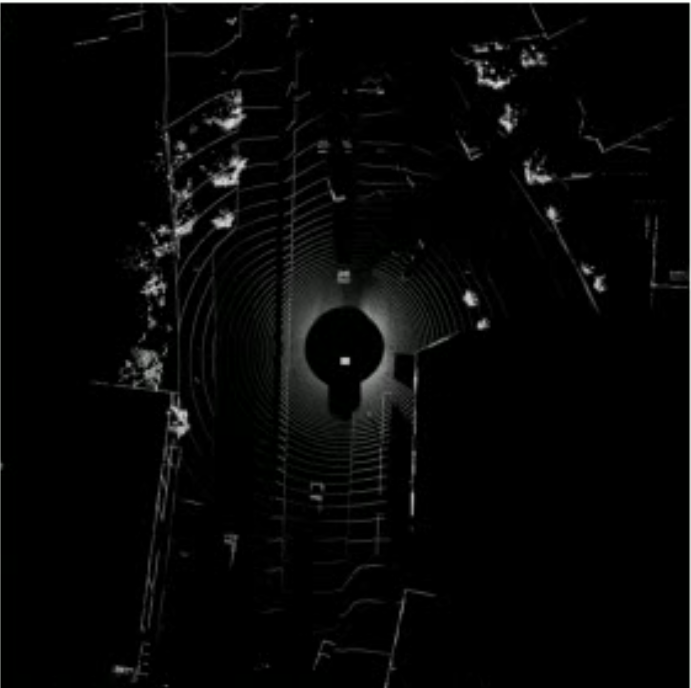}
        \caption{BEV representation of the point cloud generated by the LiDAR sensor attached to a vehicle.}
        \label{fig:lidar_image}
    \end{subfigure}
    
    \caption{Comparison of the CARLA high-fidelity and microscopic SUMO traffic simulations, along with sensor perspectives including RGB camera and LiDAR outputs from a simulated vehicle.}
    \label{fig:co-simulation}
\end{figure}
By employing the 3D raytracing detection modeling, a more realistic estimation can be made. However, the method assumes that we can detect all traffic participants that are classified as detectable in the KITTI dataset, based on the corresponding thresholds. In reality, current 3D detection algorithms can not detect all traffic participants with the given sensor information. Consequently, even with 3D raytracing, there is an overestimation of which vehicles can actually be detected and thus made available to the ITS algorithms. To mitigate this issue, the abstract microscopic traffic simulation must be enhanced to create a more realistic environment, which also allows for the simulation of realistic sensor behavior to consider the effectiveness of the 3D detection algorithms. Simulations for the training and evaluation of automated driving functions can address these challenges, as they require environments that closely resemble reality and enable accurate sensor simulations. Among the most widely used high-fidelity simulation platforms is CARLA \cite{CARLA2017}, which leverages Unreal Engine game engine to provide realistic and versatile environments for autonomous driving research. CARLA provides the capability to utilize pre-defined traffic networks, including their three-dimensional environments, or to generate custom environments based on three-dimensional scans of real traffic areas. For instance, Figure \ref{fig:co-simulation} illustrates a custom-generated map that includes three intersections from Ingolstadt, Germany. These maps allow for the definition and inclusion of streets and other traffic-relevant objects, such as traffic lights. Within these traffic networks, vehicles can be spawned and controlled, enabling the simulation of complex traffic scenarios. Furthermore, CARLA allows for the attachment of various sensors to vehicles, each configured with specific intrinsic and extrinsic parameters. Those sensors include commonly used RGB cameras, LiDARs, and RADARs, which are designed to emulate the behavior of their real-world counterparts, enabling 3D object detectors trained in the synthetic environment to achieve performance comparable to that in real-world scenarios. Hence, the CARLA simulator potentially allows for further narrowing of the gap between $D_t$ of real-world FCO systems and simulated ones. Another advantage of CARLA is its ability to integrate with the SUMO microscopic traffic simulation through a co-simulation setup. In this configuration, traffic participants and the control of traffic signals are managed by the SUMO simulation, leveraging its capabilities to easily generate specific traffic scenarios and utilize established training and evaluation pipelines for ITS systems. Meanwhile, in the CARLA simulation, specific sensor setups can be attached to vehicles in the traffic system, providing raw sensor data of the sensors of the FCO vehicles currently in the simulation according to the current penetration rate $p$. The interaction between SUMO and CARLA, as well as a visualization of the sensor data, are presented in Figure \ref{fig:main}. 
\newline
\newline
As described in Section \ref{sec:background_detection}, state-of-the-art 3D object detection algorithms utilize raw sensor information to detect and localize traffic participants captured by the sensors. These models can also be applied to synthetic sensor data the CARLA simulator generates. However, despite the high-fidelity of the simulation, a domain gap persists between data from real sensors and synthetic sensors in the simulation. Consequently, the performance of 3D object detectors -trained on real-world datasets- on synthetic data is typically lower than on real-world data. Synthetic datasets must be generated using the simulation environment to address this issue. These datasets can then be used to train existing algorithms, allowing them to adapt to the characteristics of synthetic data. Specifically, we leverage the previously introduced KITTI dataset as a benchmark. To replicate the dataset within CARLA, we place sensors on a vehicle with the same configuration as in the original KITTI dataset. This setup includes a Velodyne HDL-64E LiDAR sensor and stereo RGB cameras with a resolution of $1242 \times 375$ pixels. While the real KITTI dataset features a single forward-facing stereo RGB camera, we extend this configuration for synthetic dataset generation to a four-stereo-RGB camera setup, providing a 360° view around the vehicle to better reflect modern vehicle designs. As the RGB camera intrinsics remain consistent across all RGB cameras and each contributes independently to the dataset, this extension aligns with the single-RGB camera setup without altering the overall data processing methodology. With the sensor setup attached to several vehicles, we run the simulation and hereby collect and evaluate the generated sensor data. The evaluation hereby includes labeling of the other vehicles based on the KITTI criteria. While for the real-world KITTI dataset, this labeling is done by human annotators, within the simulated environment, we can utilize an automatic labeling pipeline based on \cite{nozarian2024CARLAkitti}. The generated labels and sensor data are stored in a structure and format consistent with the original KITTI dataset, ensuring compatibility with established algorithms and tools. A total of 16,000 individual data points are generated across five maps. Data points from the four maps provided by CARLA are used for training, while those generated from a digital twin of Ingolstadt, Germany, serve as the test set. This separation enables a robust evaluation, effectively highlighting the trained networks' ability to generalize to unseen environments.
\newline
\newline
Using the generated synthetic dataset, structured to mirror the KITTI format, we enable the training of numerous state-of-the-art 3D object detection algorithms, which are designed to accommodate this specific data structure. These algorithms can leverage RGB camera data, LiDAR data, or a combination of both to detect and localize surrounding vehicles. In this work, we demonstrate this capability using two distinct architectures: the RGB camera-based MonoCon \cite{Liu_Xue_Wu_2022} and the LiDAR-based PointPillars \cite{lang2019pointpillars}. MonoCon enhances monocular, i.e. single RGB camera, 3D object detection by introducing auxiliary tasks that learn contextual information, such as projected 2D supervision signals like key points, offset vectors, and bounding box dimensions. These auxiliary tasks are used only during training to improve detection accuracy and are discarded during inference. This design enables MonoCon to achieve competitive accuracy and fast inference times on the KITTI benchmark. PointPillars, on the other hand, is a lightweight and efficient LiDAR-based 3D object detection architecture. It organizes point cloud data into vertical columns, converting sparse point clouds into a pseudo-image representation. This approach enables the use of CNNs for feature extraction, significantly improving computational efficiency. Despite its simplicity, PointPillars delivers state-of-the-art detection performance with low latency. PointPillars achieves a BEV AP@70 of 79.83 on the "hard" split of the KITTI dataset, whereas MonoCon attains a BEV AP@70 of 19.00. This notable disparity underscores the LiDAR-provided depth information for 3D object detection tasks. LiDAR sensors deliver precise spatial depth data, facilitating accurate object localization and differentiation in three-dimensional space. In contrast, RGB-based methods like MonoCon rely predominantly on visual features, which are less effective in capturing depth and therefore achieve lower performances on the 3D detection task.
\newline
\newline
For training on the synthetic KITTI-like dataset generated in CARLA, we utilize the architectures pre-trained on the original dataset and perform fine-tuning. Fine-tuning is conducted until an average precision (AP) comparable to that of the real dataset is achieved. Although both networks have the potential to attain better performance on the synthetic dataset, we terminate training early to better approximate the performance of real-world systems within the simulation. Specifically, this entails stopping the training of the PointPillar network after 25 epochs and the MonoCon network after 175 epochs. On the test split of the synthetic dataset, the networks achieve an AP@70 of 82.24 and 20.64, respectively. In contrast, directly applying the non-fine-tuned models to the test dataset results in an AP of 15.2 and 11.0, demonstrating that the fine-tuning process effectively bridges the domain gap. With the trained 3D detectors, bounding boxes can be predicted within the CARLA simulation environment. However, instead of predicting 3D bounding boxes from the trained 3D detection algorithm, we aim to determine a binary detectability label for each vehicle to effectively use the detections within the microscopic simulation. To achieve this, we follow the intuition behind the AP metric described in \ref{sec:background_detection}, used to evaluate the performance of 3D detection algorithms. Specifically, we consider the overlap between predictions and ground-truth annotations, using the BEV IoU calculation. This approach reduces the computational complexity compared to the 3D IoU. During inference, since we lack direct associations between predicted bounding boxes and actual vehicles, we compute the IoU of each predicted box against all ground-truth instances. Any vehicle that achieves an IoU above a predefined threshold \(\tau\) is considered detected. Formally:
\[
D = \{ V_i~\in~V \mid \exists B_j~\in~\mathcal{B} : \frac{\mathrm{area}(V_i \cap B_j)}{\mathrm{area}(V_i \cup B_j)} > \tau \}\,.
\]
Using the co-simulation setup, which establishes a mapping between vehicles in the CARLA simulation and their corresponding counterparts in the SUMO simulation, we can effectively align detections from the CARLA simulation with those within the SUMO simulation.
\begin{figure}[ht]
    \centering
    \includegraphics[width=\textwidth]{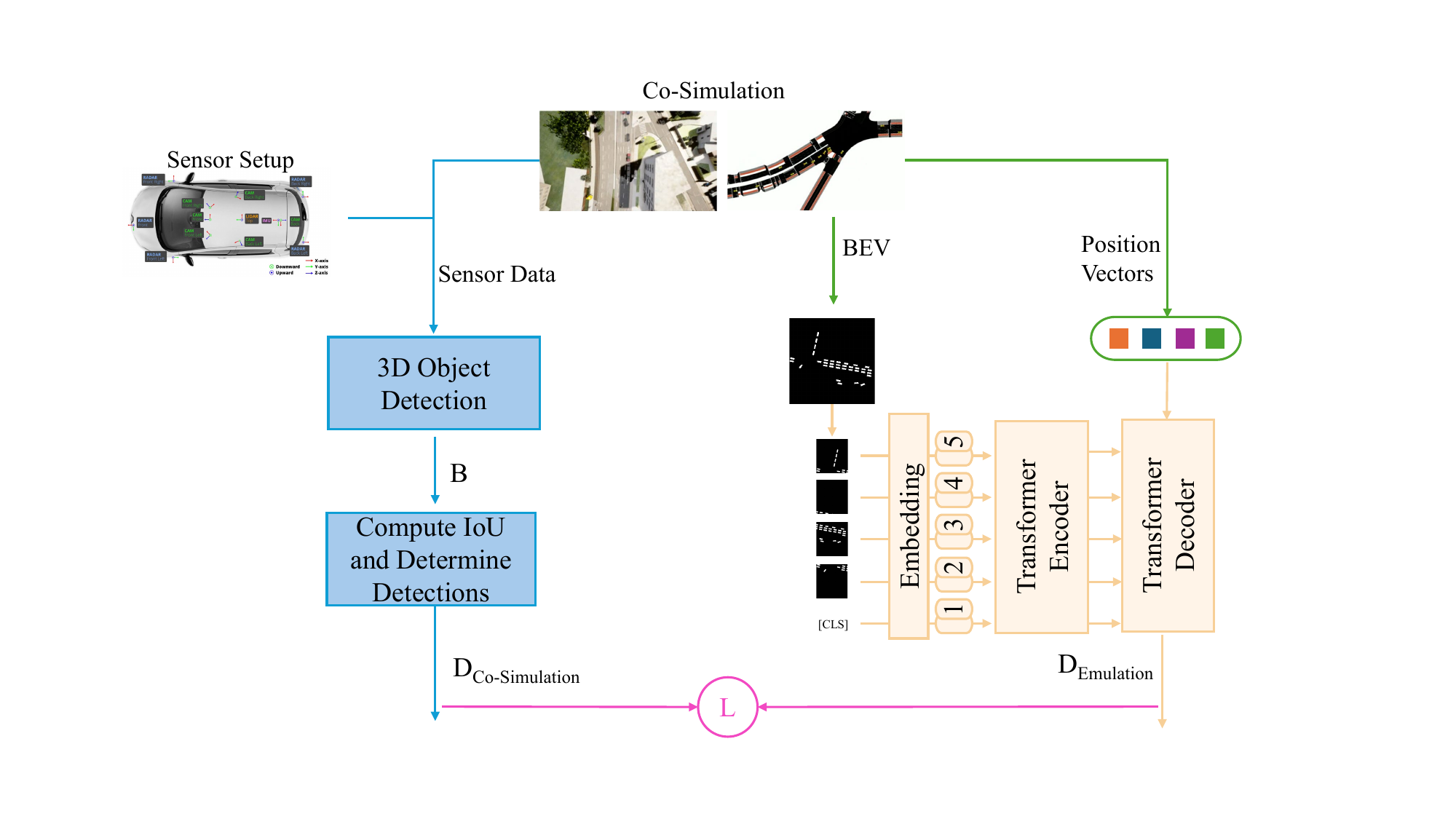}
    \caption{Illustration of the interaction between CARLA (blue), SUMO (green), and the network architecture of the emulation approach (orange) for training the emulation method. Further, the components for the loss calculation to train the emulation network are visualized (pink).}
    \label{fig:emulation_process}
\end{figure}
\subsection{Emulation of Detection Capabilites}\label{subsec:emulation}
The integration of high-fidelity traffic simulations in co-simulation frameworks, alongside microscopic traffic simulations, allows for detailed modeling of detection and system dynamics crucial for the development and evaluation of ITS. However, this approach entails significant overhead. Specifically, it requires the generation of high-resolution traffic maps for the area under investigation, tailored to test ITS systems that utilize FCO data. Furthermore, high-fidelity simulations demand specialized hardware, e.g. GPUs with minimum requirements. Despite the availability of substantial computational resources, the scalability of high-fidelity simulations is limited. As the number of FCOs and corresponding sensors in the system increases, simulation performance deteriorates, leading to reduced execution speeds. Consequently, while the co-simulation setup provides valuable insights, its scalability limitations make it impractical for large-scale analyses of ITS.
\newline
\newline
A potential solution to address this challenge is the concept of emulating detection systems using neural networks, leveraging only the information available in the microscopic traffic simulation during inference. An initial implementation of such a system was introduced in \cite{gerner2023sumodetector}. In this approach, BEV representations generated from the SUMO simulation were utilized to emulate the 3D ray-tracing method described in \ref{subsec:3D_raytracing}. These BEV representations are centered around the currently observed FCO and include nearby traffic participants as well as BEV renderings of surrounding buildings and other objects. The goal is to predict a binary classification label for each vehicle, indicating whether the respective vehicle is detectable or not. To achieve this, the system uses the BEV image alongside a position vector for each traffic participant within the BEV. Consequently, the neural network receives as input an image-like tensor representing the BEV~$\in~\mathbb{R}^{C, W, H}$ (where $C$ is the number of channels, $W$ is the width, and $H$ is the height of the image) and a position vector $p~\in~\mathbb{R}^2$, and predicts the detection outcome $d~\in~\{0, 1\}$. The model is trained on a dataset that includes, in addition to the BEV and position data, the corresponding detection labels derived from 3D ray-tracing simulations. These labels serve as ground truth for training, enabling the network to learn the mapping between BEV representations, positional information, and detection outcomes.
\newline 
\newline
Similarly, using the co-simulation between SUMO and CARLA, we can construct a dataset where the detection labels are derived from the results of the high-fidelity simulation. Such a dataset can encompass the configuration of a sensor setup and 3D detection algorithm. As visualized in Figure \ref{fig:emulation_process} utilizing the co-simulation and a specific sensor setup a sensor data stream is generated by the vehicles equipped with the sensor setup. Using a trained 3D-Object detector, bounding boxes $\mathcal{B}$ are generated from the sensor stream. In this work, we utilize the sensor setup described in Section \ref{subsec:co-simulation}, employing both the LiDAR-based PointPillar detector and the camera-based MonoCon algorithm. As outlined in Section \ref{subsec:co-simulation}, binarization is applied to match the bounding boxes with vehicles in the simulation, effectively identifying the detected vehicles in the co-simulation framework, denoted as $D_{\text{Co-simulation}}$. In parallel, for each vehicle transmitting sensor data, a Bird's Eye View (BEV) representation of the traffic situation is generated and stored at each time step, along with position vectors for nearby vehicles relative to the observed vehicle.
\newline
\newline
For the dataset, we utilize the co-simulation setup to generate training data from the default CARLA towns and test data from representations of intersections in Ingolstadt, Germany. During the simulations, for several timesteps, we collect sensor data and process it using the trained 3D detection algorithms to obtain predicted 3D bounding boxes. Detection thresholds, as described in Section \ref{subsec:co-simulation}, are applied to identify which vehicles are detectable by the respective FCO. In addition to the positional information vectors and BEV data, we generate data points to train and evaluate the emulation method. Unlike the initial approach, which stored BEV images directly from the SUMO simulation, we instead save only the raw polygonal information of traffic participants and surrounding objects. This approach enables a more streamlined generation of the input tensor representing the BEV for the neural networks during training and inference. We extract polygonal information of buildings and other environmental objects from the high-fidelity simulation using a LiDAR sensor mounted on a vehicle within the CARLA simulation environment. The vehicle traverses all streets across the various maps, generating a point cloud that includes objects potentially causing occlusions during 3D object detection. To focus on relevant occlusions, we restrict the point cloud to a 2.5-meter range from the street, as points beyond this distance are unlikely to contribute to occlusions. These points are then projected onto the ground plane and clustered using the DB-SCAN algorithm \cite{DBSCAN}. Subsequently, convex hulls are constructed around each cluster to generate polygons representing potential occluding objects. The corresponding polygons for the CARLA map of Ingolstadt, Germany, are illustrated in Figure \ref{fig:co-simulation}. Utilizing three CARLA maps not used for the generation of the 3D object detection dataset, we generate a total of about $1 \times 10^5$ datapoints for both the MonoCon and Pointpillar detectors.
\newline
\newline
Due to the limited number of CARLA maps and, consequently, the restricted variety of traffic scenarios represented in the dataset, achieving good generalization for training the neural networks is challenging. Specifically, while the emulation of detection performs well for scenarios covered by the CARLA maps, it fails to generalize effectively to scenarios outside of this coverage. To address this problem, we mitigate the limitation by performing pre-training with a different detection method that can cover a broader range of traffic scenarios. We then perform fine-tuning using the smaller dataset from the co-simulation. This fine-tuning can be seen as a calibration step to align the model with the actual detection characteristics. We use the previously presented 2D-raytracing detection modeling to generate the pre-training dataset since it is the fastest of the detection models introduced allowing for fast generation of data points i.e. generation of large and diverse datasets. For this generation we utilize the SUMO digital twin of Ingolstadt, Germany presented in \cite{harth2021automated} which also provides realistic traffic demands. This SUMO traffic network also includes the intersection of the realistic high-fidelity map used in Section \ref{subsec:co-simulation}. We specifically collect data points for vehicles within a range of 125 meters from 40 randomly selected intersections during the simulation. This process results in a total of $3 \times 10^6$ individual data points and $2.5 \times 10^5$ corresponding BEV representations. On average, each BEV representation covers a 50-meter radius around the respective floating car object and includes approximately 12 vehicles. The dataset exhibits a positive-to-negative sample ratio of 1.35 to 1. For training and testing, we use 90\% of the intersections and the data generated by them as training data, and reserve 10\% for testing. This split ensures that the model's generalizability is evaluated across different traffic network structures. Furthermore, when creating the training and testing split, we explicitly include the intersection where we have the high-fidelity simulation into the test data.
\newline
\newline
In previous work \cite{gerner2023sumodetector}, we explored both CNN-based ResNet \cite{RESNET} architectures and transformer-based ViT \cite{ViT} architectures to model detectability from the dataset. In each case, the models utilized their respective backbones to extract dense feature representations from the input BEV image. These representations were then concatenated with a two-dimensional position vector and processed through an MLP to produce the final binary output probability. In this work, we enhance the network architecture by more effectively integrating the position vector into the decision-making process. Specifically, we implement an encoder-decoder architecture that again employs the ViT as an encoder, while incorporating the position vector within a transformer-decoder structure. Beginning with the encoder, we follow the standard ViT approach. The input image is divided into a predefined grid of patches, where each patch is flattened and linearly projected into an embedding of dimension $d$. Additionally, a learnable [CLS] token is prepended to the sequence of patch embeddings. This token represents a global summary of the entire image, accumulating contextual information from all patches. These embeddings, including the [CLS] token, are passed through the transformer encoder, which uses multiple layers of self-attention and feed-forward networks. Self-attention calculates how each token relates to every other token, allowing global context and relationships across the image to be captured efficiently. After this encoding process, the [CLS] token and the refined patch embeddings form a rich representation of the scene. Previously, and in the original formulation of the ViT for image classification, only the transformed [CLS] token was subsequently used for further processing. In contrast, in this work, we leverage all transformed tokens by feeding them into a transformer-based decoder. In this decoder, the tokens serve as the keys of the cross-attention mechanism, while the position vector, after being embedded into the same $d$-dimensional space as the encoder embeddings, provides the queries and values. Through the cross-attention layers, the model learns to relate each position embedding (query) with all encoded visual features (keys and values) extracted by the encoder. This process allows the position vector to selectively aggregate relevant spatial and contextual information from the encoded image representation. In other words, the decoder can attend to different regions of the scene based on the positional input. Finally, the refined embedding is passed through an MLP that produces the binary detectability probability, ultimately leading to more accurate and context-aware detectability predictions. This architecture is illustrated in Figure \ref{fig:emulation_process}.
\newline
\newline
To showcase the benefits of the new architecture, we train both the previous ViT-based architecture and the adapted encoder-decoder architecture using the same hyperparameters in the encoder part. Specifically, we utilize a patch size of $32 \times 32$ pixels on the BEV image, with height and width set to 256 pixels and a single channel. The embedding dimension is 512, with 4 heads in the multi-head self-attention mechanism and 3 transformer layers in the encoder. For the transformer decoder, we use 4 attention heads and 2 layers. Both architectures are trained on the previously introduced pre-training dataset, where the 2D raytracing was used as the target, for 100 epochs. The encoder-only architecture achieves a detection accuracy of 85.4\%. In contrast, our new encoder-decoder architecture achieves an accuracy of 93.6\%, clearly outperforming the previous approach. This demonstrates that integrating positional information into a transformer decoding process significantly enhances the model's ability to interpret spatial and contextual relationships, leading to improved detectability predictions.
The pre-trained encoder-decoder architecture is utilized to fine-tune each of the smaller datasets generated with the co-simulation framework. We achieve a mean accuracy of 86.6\% for the emulation of MonoCon detections and 90.7\% for the emulation of PointPillars detections on the test split of the dataset. As with the 3D object detection, we structure the test split such that the test data originates from the high-fidelity simulation of Ingolstadt, Germany. This approach demonstrates the generalizability of the method to different traffic scenarios, which is also enhanced by the pre-training process. We hypothesize that the slightly lower performance of the camera-based detection emulation can be attributed to the BEV representation, which serves as the input for the neural network. This representation does not fully capture factors such as potential shadowing effects from buildings, which can influence camera-based detection accuracy. Overall, the emulation method proves to be effective in approximating the detections of real 3D object detectors within the co-simulation setup. Notably, it requires only the information from the microscopic simulation during inference, making the co-simulation process obsolete.
\begin{table}[htb]
\centering
\begin{tabular}{|l|c|c|c|}
\hline
\diagbox{3D Detector}{Detection Model} & 2D-Raytracing & 3D-Raytracing & Emulation \\ \hline
Monocon             &                77.8\%              &       78.1\%        &      86.6\%    \\ \hline
PointPillar          &                  75.9\%            &        83.0\%       &       90.7\%  \\ \hline
\end{tabular}
\caption{Accuracy comparison of detection modeling techniques using the Co-Simulation approach, with the respective 3D detector serving as the ground truth reference. For the emulation modeling, the trained model corresponding to each 3D detector is utilized. The 2D ray-tracing method employs 360 rays and a detection threshold of one hit. The 3D raytracing approach follows detection thresholds defined by the "hard" KITTI criteria.}
\label{tab:detection_comparison}
\end{table}
\subsection{Comparison of Detection Modeling Techniques}\label{subsec:modeling_evaluation}
In the previous sections, we presented a set of methods for modeling the detection capabilities of real-world FCOs within a microscopic traffic simulation. Key quality characteristics of these models are their ability to replicate detections as accurately as possible, closely matching those observed in comparable real-world scenarios and their minimal impact on the simulation speed of the microscopic traffic model. This ensures the efficient development, training, and evaluation of ITS when using FCO information. Regarding the accuracy of detections, the co-simulation modeling approach emerges as the most precise method presented. By employing emulated sensors in high-fidelity simulators with high-fidelity maps and using 3D object detectors that are also applied in real-world scenarios, the difference between simulated and real detections is minimized. With continued advancements in high-fidelity simulators, it is expected that this domain gap will further decrease over time. For this reason, we consider the co-simulation method as the most accurate modeling approach in this study and use it as the benchmark against which other modeling methods are compared. For an evaluation of alternative modeling methods, we run the co-simulation method on the CARLA map of Ingolstadt, Germany, with a set of FCOs and gather the $D_{\text{co-simulation}}$ observers at different timesteps. In parallel, we also generate the detections for the same FCOs using 2D raytracing, 3D raytracing, and the neural network-based emulation modeling approach. Table \ref{tab:detection_comparison} summarizes the results, presenting the accuracy of each method relative to the co-simulation benchmark for both camera- and LiDAR-based detections. Overall, the 2D raytracing method exhibits the lowest accuracy, achieving only 77.8\% for MonoCon detections and 75.9\% for PointPillar detections within the co-simulation setup. A detailed examination of the detections reveals both false positives and false negatives. In some cases, the 2D raytracing method overestimates the detection capabilities of the 3D object detectors. Additionally, while the MonoCon approach generally detects fewer objects due to its lower average precision (AP), it also underestimates detections by incorrectly predicting that certain objects are not detectable when they actually are, likely due to the 2D abstraction of the method. Similarly, the 3D raytracing approach shows both over- and underestimation of individual detections. Underestimations are particularly prevalent in LiDAR-based detection, where some vehicles are detected by the algorithm despite being labeled as "DontCare" in the KITTI dataset, indicating that the "hard" thresholds are not met. The emulation modeling technique achieves the best performance, as it can overcome the limitations highlighted by the 2D raytracing through the learning process, despite its 2D abstraction. 
\newline
\newline
In addition to capturing the detection accuracy of the various modeling methods, we also record the time required for each method to generate detections. The performance evaluation is conducted on a workstation equipped with an NVIDIA A6000 GPU and an AMD Ryzen Threadripper PRO 5965WX CPU. To assess hardware dependency, additional evaluations are performed on devices with differing capabilities: a laptop lacking GPU support and equipped with an AMD Ryzen 5 CPU. These tests provide a comparative analysis of computational efficiency across varying hardware configurations, offering insights into the scalability and practical applicability of the proposed modeling techniques. The results of this analysis are presented in Table \ref{tab:detection_speed}. It is important to highlight that the co-simulation approach is only feasible on systems equipped with GPUs, as the CARLA simulation necessitates GPU resources. Even on a high-performance workstation, determining the detected vehicles for each FCO requires an average of 803 ms. This duration encompasses the overhead from the CARLA simulation, including sensor data generation, as well as the forward pass through the 3D object detectors and the assignment of detections based on the Intersection over Union (IoU) metric. In contrast, the 3D raytracing method demands even more time, taking several seconds per modeling instance. From a temporal perspective, both of these methods appear unsuitable for large-scale applications due to their significant computational demands. On the other hand, the 2D raytracing method is markedly faster, requiring an average of 102 ms or 213 ms, depending on the system configuration. The most efficient modeling method identified is the neural network-based emulation, which requires only 14.8 ms when utilizing a GPU and performing efficient forward passes through batching of FCOs. Notably, even without GPU support, the emulation method remains the fastest modeling approach, enabling large-scale FCO elaboration even on systems with limited computational resources.
\newline
\newline
\newline
\newline
\newline
\begin{table}
\centering
\begin{tabular}{|l|c|c|c|c|}
\hline
\diagbox{CPU/GPU}{Detection Model} & 2D raytracing & 3D raytracing & Co-simulation & Emulation \\ \hline
\makecell{AMD Threadripper \\ NVIDIA A6000}      &      102ms         &        6.18s       &       803ms        &     14.8ms      \\ \hline
\makecell{AMD Ryzen 5 \\ /}                      &        213ms       &       13.4s        &      /        &      83.5ms     \\ \hline
\end{tabular}
\caption{Average duration of FCO detectability evaluations for different detection modeling techniques and hardware configurations.}
\label{tab:detection_speed}
\end{table}

\clearpage
\FloatBarrier
\begin{figure}[h]
\centering
\begin{subfigure}[t]{0.3\textwidth}
    \centering
    \includegraphics[width=\textwidth]{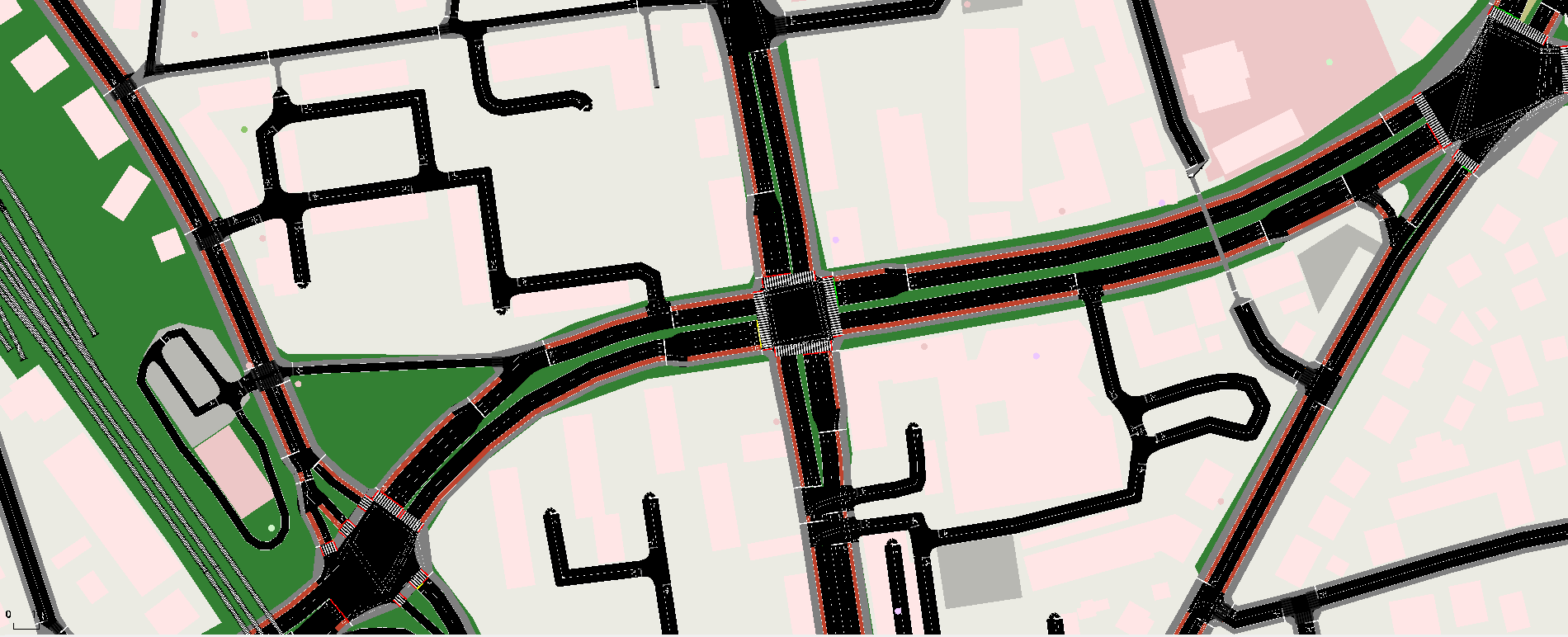} 
\end{subfigure}%
\hfill
\begin{subfigure}[t]{0.3\textwidth}
    \centering
    \includegraphics[width=\textwidth]{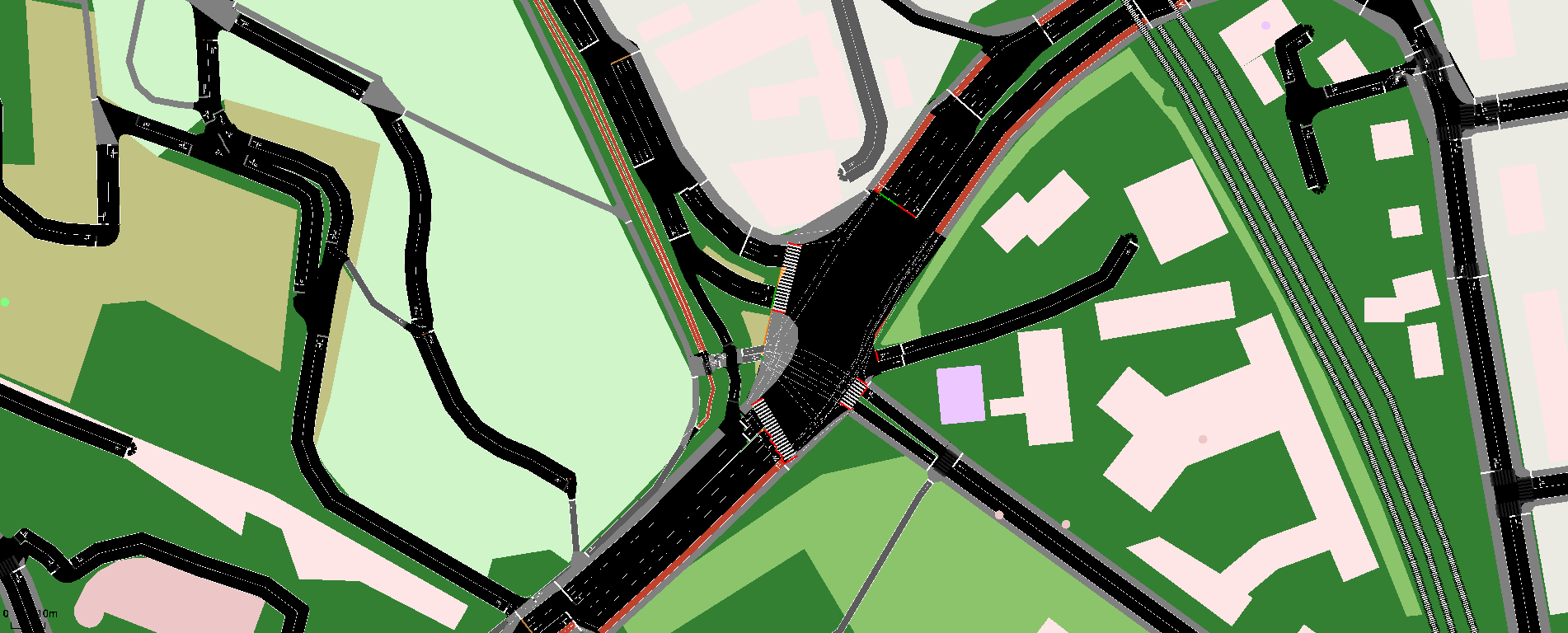}
\end{subfigure}%
\hfill
\begin{subfigure}[t]{0.3\textwidth}
    \centering
    \includegraphics[width=\textwidth]{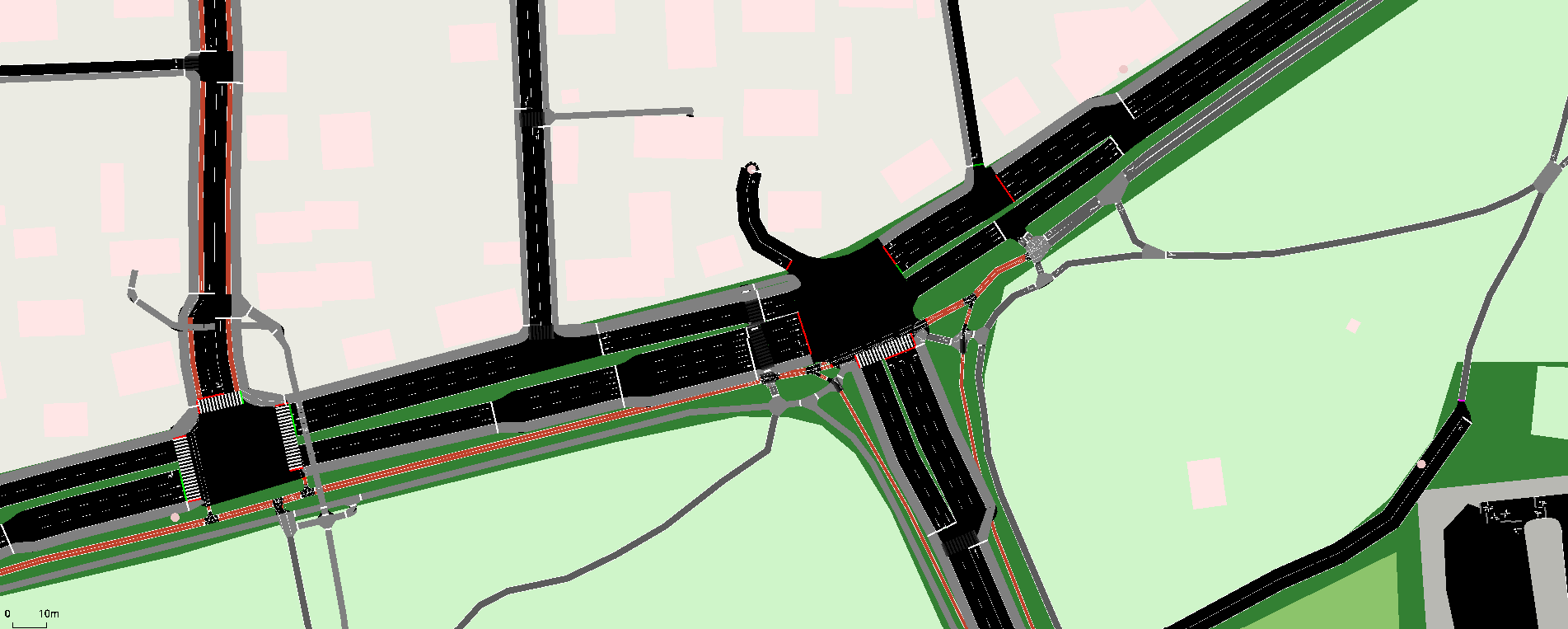}
\end{subfigure}

\caption{Excerpt of the SUMO simulation visualizing the investigated intersections for the FCO and temporal insights potential.}
\label{fig:intersection_images}
\end{figure}

\begin{table}[!ht]
\centering
\begin{subtable}[t]{0.32\textwidth} 
    \centering
    \begin{tabular}{|c|c|c|c|}
    \hline
    \diagbox{$q$}{$p$} & 5\% & 10\% & 20\% \\ 
    \hline
    low & \cellcolor{blue!11} 11\% & \cellcolor{blue!19} 19\% & \cellcolor{blue!50} 50\% \\ 
    \hline
    med & \cellcolor{blue!20} 20\% & \cellcolor{blue!30} 30\% & \cellcolor{blue!56} 56\% \\ 
    \hline
    high & \cellcolor{blue!24} 24\% & \cellcolor{blue!44} 44\% & \cellcolor{blue!58} 58\% \\ 
    \hline
    \end{tabular}
\end{subtable}%
\hfill 
\begin{subtable}[t]{0.32\textwidth}
    \centering
    \begin{tabular}{|c|c|c|c|}
    \hline
    \diagbox{$s$}{$p$} & 5\% & 10\% & 20\% \\ 
    \hline
    low & \cellcolor{blue!9} 9\% & \cellcolor{blue!22} 22\% & \cellcolor{blue!50} 50\% \\ 
    \hline
    med & \cellcolor{blue!13} 13\% & \cellcolor{blue!34} 34\% & \cellcolor{blue!57} 57\% \\ 
    \hline
    high & \cellcolor{blue!9} 9\% & \cellcolor{blue!28} 28\% & \cellcolor{blue!54} 54\% \\ 
    \hline
    \end{tabular}
\end{subtable}%
\hfill
\begin{subtable}[t]{0.32\textwidth}
    \centering
    \begin{tabular}{|c|c|c|c|}
    \hline
    \diagbox{$q$}{$p$} & 5\% & 10\% & 20\% \\ 
    \hline
    low & \cellcolor{blue!21} 21\% & \cellcolor{blue!26} 26\% & \cellcolor{blue!57} 57\% \\ 
    \hline
    med & \cellcolor{blue!23} 23\% & \cellcolor{blue!41} 41\% & \cellcolor{blue!67} 67\% \\ 
    \hline
    high & \cellcolor{blue!28} 28\% & \cellcolor{blue!35} 35\% & \cellcolor{blue!63} 63\% \\ 
    \hline
    \end{tabular}
\end{subtable}%
\caption{FCO potential utilizing the emulation modeling trained on the camera-based MonoCon detector at the different penetration rates $p$ and traffic demands $q$ for the respective intersection shown above.}
\label{tab:fco_potential_monocon}
\end{table}

\begin{table}[ht!]
\centering
\begin{subtable}[t]{0.32\textwidth} 
    \centering
    \begin{tabular}{|c|c|c|c|}
    \hline
    \diagbox{$q$}{$p$} & 5\% & 10\% & 20\% \\ 
    \hline
    low & \cellcolor{blue!32} 32\% & \cellcolor{blue!51} 51\% & \cellcolor{blue!71} 71\% \\ 
    \hline
    med & \cellcolor{blue!36} 36\% & \cellcolor{blue!57} 57\% & \cellcolor{blue!76} 76\% \\ 
    \hline
    high & \cellcolor{blue!26} 26\% & \cellcolor{blue!58} 58\% & \cellcolor{blue!79} 79\% \\ 
    \hline
    \end{tabular}
\end{subtable}%
\hfill 
\begin{subtable}[t]{0.32\textwidth}
    \centering
    \begin{tabular}{|c|c|c|c|}
    \hline
    \diagbox{$s$}{$p$} & 5\% & 10\% & 20\% \\ 
    \hline
    low & \cellcolor{blue!29} 29\% & \cellcolor{blue!51} 51\% & \cellcolor{blue!70} 70\% \\ 
    \hline
    med & \cellcolor{blue!21} 21\% & \cellcolor{blue!38} 38\% & \cellcolor{blue!72} 72\% \\ 
    \hline
    high & \cellcolor{blue!31} 31\% & \cellcolor{blue!52} 52\% & \cellcolor{blue!64} 64\% \\ 
    \hline
    \end{tabular}
\end{subtable}%
\hfill
\begin{subtable}[t]{0.32\textwidth}
    \centering
    \begin{tabular}{|c|c|c|c|}
    \hline
    \diagbox{$q$}{$p$} & 5\% & 10\% & 20\% \\ 
    \hline
    low & \cellcolor{blue!21} 21\% & \cellcolor{blue!37} 37\% & \cellcolor{blue!66} 65\% \\ 
    \hline
    med & \cellcolor{blue!34} 33\% & \cellcolor{blue!54} 53\% & \cellcolor{blue!83} 83\% \\ 
    \hline
    high & \cellcolor{blue!26} 25\% & \cellcolor{blue!31} 30\% & \cellcolor{blue!68} 67\% \\ 
    \hline
    \end{tabular}
\end{subtable}%
\caption{FCO potential utilizing the emulation modeling trained on the LiDAR-based Pointpillar detector at the different penetration rates $p$ and traffic demands $q$ for the respective intersection shown above.}
\label{tab:fco_potential_pointpillar}
\end{table}

\begin{table}[ht!]
\centering
\begin{subtable}[t]{0.3\textwidth}
    \centering
    \begin{tabular}{|c|c|c|c|}
    \hline
    \diagbox{$s$}{$p$} & 5\% & 10\% & 20\% \\ 
    \hline
    5 & \cellcolor{purple!10} 9.6\% & \cellcolor{purple!9} 9.4\% & \cellcolor{purple!7} 7.2\% \\ 
    \hline
    10 & \cellcolor{purple!18} 18\% & \cellcolor{purple!16} 16\% & \cellcolor{purple!12} 12\% \\ 
    \hline
    20 & \cellcolor{purple!29} 29\% & \cellcolor{purple!28} 28\% & \cellcolor{purple!16} 16\% \\ 
    \hline
    \end{tabular}
\end{subtable}%
\hfill
\begin{subtable}[t]{0.3\textwidth}
    \centering
    \begin{tabular}{|c|c|c|c|}
    \hline
    \diagbox{$s$}{$p$} & 5\% & 10\% & 20\% \\ 
    \hline
    5 & \cellcolor{purple!5} 5.3\% & \cellcolor{purple!9} 8.8\% & \cellcolor{purple!9} 8.8\% \\ 
    \hline
    10 & \cellcolor{purple!10} 9.9\% & \cellcolor{purple!16} 16\% & \cellcolor{purple!14} 14\% \\ 
    \hline
    20 & \cellcolor{purple!17} 17.4\% & \cellcolor{purple!25} 25\% & \cellcolor{purple!20} 20\% \\ 
    \hline
    \end{tabular}
\end{subtable}%
\hfill
\begin{subtable}[t]{0.32\textwidth}
    \centering
    \begin{tabular}{|c|c|c|c|}
    \hline
    \diagbox{$s$}{$p$} & 5\% & 10\% & 20\% \\ 
    \hline
    5 & \cellcolor{purple!6} 5.9\% & \cellcolor{purple!9} 8.6\% & \cellcolor{purple!4} 4.3\% \\ 
    \hline
    10 & \cellcolor{purple!11} 11\% & \cellcolor{purple!15} 15\% & \cellcolor{purple!7} 7.1\% \\ 
    \hline
    20 & \cellcolor{purple!18} 18\% & \cellcolor{purple!24} 24\% & \cellcolor{purple!10} 9.6\% \\ 
    \hline
    \end{tabular}
\end{subtable}%

\caption{Additional potential through FCOs when previously seen vehicles can be recovered from sequence lengths $s$ at different penetration rates $p$. Results are obtained using the emulation modeling, specifically the one that is trained on the PointPillar detector.}
\label{tab:tfco_potential}
\end{table}
\FloatBarrier

\subsection{Comparing Floating Car Data to the extended Floating Car Data}
Modeling detection capabilities within a microscopic traffic simulation allows to assess the additional information provided by the data source compared to standard Floating Car Data (FCD) and to evaluate its potential benefits for ITS. This analysis is conducted using the SUMO simulation of Ingolstadt, Germany, which was calibrated with real traffic data and was presented by \cite{harth2021automated}. Specifically, we select five intersections with varying geometries and analyze them under three traffic demand scenarios: low (6:00 AM), medium (10:00 AM), and high (4:00 PM).\footnote{The determination of low, medium, and high demand is done by checking the total number vehicles within the simulation. I.e the interpretation of low, medium, and high is relative to the traffic that occurs during the 24h simulation.} For each scenario, we evaluate different FCD penetration rates of 5\%, 10\%, and 20\%. Based on the results in Section~\ref{subsec:modeling_evaluation}, we utilize the emulation modeling approach outlined in Section~\ref{subsec:emulation} and investigate the performance of both a camera-based detection and LiDAR-based detection given the sensor setup introduced in Section~\ref{subsec:co-simulation}.  Figure~\ref{fig:intersection_images} illustrates the selected intersections, while Table~\ref{tab:fco_potential_monocon} presents the results for the different penetration rates across the various intersections for the MonoCon-based emulation and Table~\ref{tab:fco_potential_pointpillar} shows the same for the LiDAR-based detection emulation. The presented data represents the ratio of detected vehicles and observers within the area of interest to the total number of vehicles in the same area. The results highlight the potential advantages of FCOs compared to conventional FCD information. For instance, with an FCO penetration rate of 20\% at the third analyzed intersection, a potential of 83\% is observed. This indicates that 63\% of the vehicles could be detected solely through FCO-based observations, serving as valuable traffic information. Furthermore, the results confirm the superior 3D detection performance of the PointPillars model compared to the MonoCon detections, which is effectively reflected in the emulation process. The LiDAR-based detection consistently demonstrated higher potential compared to scenarios relying solely on camera data. However, the specific potentials are inherently dependent on the traffic situation and the spatial distribution of FCOs within the traffic network.

\section{Including Temporal Insights from Floating Car Observers}
Investigations into the additional information provided by FCO compared to conventional FCD have revealed substantial potential benefits, however, there are even more advantages. In our previous work \cite{IV2024}, we highlighted that FCOs not only detect vehicles at the current timestep but also leverage past observations of vehicles that are no longer visible at the present moment. By integrating this historical information, FCOs provide additional information on the current traffic state. Formally, let \( V_{s,t} \) denote the set of vehicles detected at the current timestep \( t \) or within the preceding \( s \) steps:
\[
V_{s,t} = \left( \bigcup_{k=t-s}^{t} V_{d,k} \right) \cap V_t,
\]
where \( V_{d,k} \) is the set of detected vehicles at timestep \( k \), and \( V_t \) is the set of all vehicles present at \( t \). Since \(|V_{s,t}| \geq |V_{d,t}|\), the set \( V_{s,t} \) captures a broader temporal context. The value of this expanded temporal information depends on factors such as the FCO penetration rate, sensor configurations, the structure of the road network, the prevailing traffic conditions, and the chosen sequence length \( s \). Table \ref{tab:tfco_potential} shows the temporal potential for different sequence length for the data which was analyzed in Table \ref{tab:fco_potential_pointpillar}. It becomes evident that even with short sequence lengths, a significant number of additional vehicles can be detected if the temporal potential is utilized. This potential increases further as the sequence length is extended.
\newline
\newline
Developing neural network architectures that can recover the current positions of previously observed but now undetected vehicles requires datasets that represent diverse scenarios. Following \cite{IV2024}, we concentrate on a single intersection in Ingolstadt, Germany, within a 100-meter radius. Restricting the data to this specific region enables the model to implicitly learn the local street structure without requiring explicit network information, as is necessary for datasets like \cite{sun2020scalability}. To create these datasets, we simulate 1000-second scenarios with a timestep of \(\Delta t = 1\) second, varying FCO penetration rates (5\%, 10\%, 20\%) and start times to capture different traffic demands. At each timestep, we apply the detection emulation modeling to determine which vehicles \( D_t \) would be detected by FCOs and record their positions together with the positions of all vehicles in \( V_t \). By normalizing the positions of vehicles relative to the area of interest, we ensure consistent scaling and alignment of spatial data, reducing variability and improving the efficiency and effectiveness of neural network training later when using the dataset. The resulting dataset has a size of $3 \times e^{5}$ individual data points. We then split this dataset into training and testing subsets, ensuring that test scenarios—defined by unique combinations of start times and penetration rates—remain unseen during training. From this dataset, we extract historical sequences of length \( s \) ending at a given timestep \( t \). Vehicles that are not visible at time \( t \) may have appeared one or more times in the preceding timesteps. We define \(\alpha\) as the number of times such a previously observed, now undetected vehicle has appeared in the sequence. These sequences, along with their associated \(\alpha\) values, serve as inputs to the neural network architectures, enabling them to learn from both the immediate past and the historical presence patterns of vehicles. By incorporating this temporal perspective, the model aims to accurately recover the positions of these previously observed but currently unseen vehicles.
\newline
\newline

In our previous work \cite{IV2024}, we employed an image-to-time pipeline using a BEV representation of traffic scenes for recovering currently unseen vehicles. By converting spatial layouts into temporal sequences, this approach aimed to predict future vehicle positions from past observations. To process BEV sequences, we leveraged architectures inspired by video understanding, including 3D CNNs and Video Vision Transformers (ViViT), which jointly capture spatial and temporal dynamics. These models produced pixel-wise probability maps of vehicle positions, the best-performing model in \cite{IV2024} accurately recovered about 40\% of vehicles at the current timestep, which still has potential for improvement. We attribute current limitations to the loss functions such as IoU and weighted binary cross-entropy that are sensitive to misalignment and hinder model convergence. For tasks where a detailed probabilistic representation of the predicted vehicle positions is not essential, directly predicting positions can simplify the approach. Avoiding the BEV-based encoders reduces complexity and computational overhead, potentially improving scalability and performance. Thus, we propose a streamlined network architecture that directly consumes raw vehicle data at each timestep. By eliminating image-based feature extraction, we concentrate on learning temporal dynamics and directly predicting previously unseen vehicles’ positions.
\newline
\newline
The proposed architecture is built upon a transformer-encoder framework and leverages raw data provided by the FCOs. Specifically, the position information of all vehicles is embedded into a representation in \(\mathbb{R}^d\).
Additionally, we include a binary indicator specifying whether each vehicle is detected at the current time step. Undetected vehicles lack positional information and are explicitly marked as not detected. As a result, the embedding function maps from $\mathbb{R}^3$ to $\mathbb{R}^d$. Furthermore, we define a maximum number of vehicles, denoted as $v_{\text{max}}$, essential for batching during transformer training. This results in a combined representation of the system in $\mathbb{R}^{s \times v_{\text{max}}}$, where $s$ represents the sequence length. Instead of processing each time step individually, we flatten the embedding information into a single sequence. In other words, the input does not have the shape $\mathbb{R}^{s \times v_{\text{max}}}$. By doing this, we avoid processing vehicles sequentially within a single time step and across different time steps separately. Instead, we allow the transformer architecture to simultaneously attend to all the information. Since self-attention does not inherently capture positional information, we introduce a learnable positional encoding. This encoding provides information about the time step within the sequence $s$ that each embedding belongs to. However, this approach introduces potential challenges in the computational complexity. Flattening the embedding into a single sequence increases the input length from $s \times v_{\text{max}}$ to $s \cdot v_{\text{max}}$. Since the computational complexity of the self-attention mechanism is $O(n^2)$, where $n$ is the total sequence length, this increases the computational complexity significantly. To mitigate this challenge, masking can be employed within the self-attention mechanism. By applying a mask, we can selectively focus our attention on relevant embeddings. 
We specifically apply this embedding, such that a vehicle embedding attends to all other embeddings in the same time step and the same embedding of the same vehicle in the other timestep. Doing so, we reduce the total sequence length to $n=\mathbb{R}^{s + v_{\text{max}}}$ and thereby also the number of attentions operated. After processing the input through the transformer encoder architecture, the network produces a sequence of transformed embeddings, each corresponding to a particular timestep. To generate predictions for the current timestep, the embeddings associated with the final timestep are extracted and passed through an MLP head. Each resulting output vector \(r~\in~\mathbb{R}^{3}\) is given by
\[
r = \begin{bmatrix} r^{c} \\ r^{p} \end{bmatrix},
\]
where \(r^{c}~\in~\mathbb{R}\) is a scalar classification logit and \(r^{p}~\in~\mathbb{R}^{2}\) is the predicted position of the corresponding vehicle.
For supervised training, the target vector \(\hat{r}~\in~\mathbb{R}^{3}\) is defined as
\[
\hat{r} = \begin{bmatrix} \hat{r}^{c} \\ \hat{r}^{p} \end{bmatrix}.
\]
The component \(\hat{r}^{c}\) is always equal to one, which represents the binary classification target, and \(\hat{r}^{p}~\in~\mathbb{R}^{2}\) denotes the true vehicle position.
When computing the loss, the focus is on vehicles that were present in the input sequence but are not directly observed at the current timestep. This encourages the model to infer the current positions of vehicles that have moved out of immediate view and thus tests the network's capacity to leverage historical context.
The training objective combines a binary classification loss \(\mathcal{L}_{\text{class}}\) and a position regression loss \(\mathcal{L}_{\text{position}}\). The classification loss is defined using a binary cross-entropy criterion:
\[
\mathcal{L}_{\text{class}} = - \left[ \hat{r}^{c} \cdot \log(\sigma(r^{c})) + (1 - \hat{r}^{c}) \cdot \log(1 - \sigma(r^{c})) \right].
\]
Here \(\sigma\) maps the classification logit to a probability value between zero and one. Although the classification target is always one, this formulation allows the model to express uncertainty by producing a classification logit that is less than one in challenging scenarios.
The position loss measures the discrepancy between the predicted and the true vehicle positions:
\[
\mathcal{L}_{\text{position}} = \sigma(r^{c}) \cdot \|r^{p} - \hat{r}^{p}\|_{2}.
\]
In this formulation, \(\sigma(r^{c})\) serves as a confidence weight. It ensures the model focuses on positional accuracy primarily when it is confident in its classification estimate. When the classification output is low, the positional error penalty is effectively reduced, allowing the network to express uncertainty rather than forcing inaccurate position predictions.
A balance between classification and positional objectives is achieved by combining these losses into a single loss function:
\[
\mathcal{L} = \lambda_{\text{class}} \mathcal{L}_{\text{class}} + \lambda_{\text{position}} \mathcal{L}_{\text{position}}.
\]
The factors \(\lambda_{\text{class}}\) and \(\lambda_{\text{position}}\) determine the relative importance of classification and positional accuracy. This formulation encourages the model to produce accurate positions while maintaining an internal measure of confidence. Even when the model struggles to pinpoint the exact location of a vehicle, it remains incentivized to attempt a prediction since the classification target is always one. This creates a balanced framework that promotes both accuracy and confidence in position estimates, even for vehicles that are no longer within the immediate field of view at the current timestep.
\newline
\newline
We train the proposed architecture using the described loss function on the presented dataset. The model uses an embedding dimension \( d = 64 \), \( h = 8 \) attention heads, and \( L = 4 \) self-attention layers. We employ a cosine annealing learning rate schedule with a warm-up phase to ensure stable training. Training is conducted for 100 epochs, varying the sequence length \( s \) and additional parameters to assess their impact. We investigate the effect of \(\alpha_{\text{min}}\), which determines the minimum number of times a vehicle must appear in the sequence to be included as a recovery target. Higher \(\alpha_{\text{min}}\) reduces the number of recoverable vehicles, while lower values may challenge the model with sparse historical information. Furthermore, we evaluate different values of \(\lambda_{\text{position}}\) in the loss function while keeping \(\lambda_{\text{class}}\) constant. As evaluation criteria, we use two metrics to assess the performance of the proposed approach. The first metric is the recovery accuracy \(a\), which measures the proportion of recoverable vehicles, defined by \(\alpha_{\text{min}}\), that are successfully recovered. A vehicle is considered successfully recovered if the classification score \(r^{(c)}\), after applying the sigmoid function \(\sigma(\cdot)\), exceeds a threshold of 0.5. Formally, this is defined as:
\[
a = \frac{\sum_{i=1}^{N} \mathbb{1}\left[\sigma\left(r^{(c)}_i\right) > 0.5\right]}{\sum_{i=1}^{N} \mathbb{1}\left[\hat{r}^{(c)}_i = 1\right]},
\]
where \(\mathbb{1}[\cdot]\) is the indicator function, \(N\) is the total number of vehicles in the dataset, and \(\hat{r}^{(c)} = 1\) identifies recoverable vehicles. The second metric is the mean Euclidean distance $\overline{e}$ between the predicted position \(r^{(p)}\) and the target position \(\hat{r}^{(p)}\), calculated only for vehicles where \(\sigma\left(r^{(c)}\right) > 0.5\). This is expressed as:
\[
\overline{e} = \frac{\sum_{i=1}^{N} \mathbb{1}\left[\sigma\left(r^{(c)}_i\right) > 0.5\right] \cdot \|r^{(p)}_i - \hat{r}^{(p)}_i\|_2}{\sum_{i=1}^{N} \mathbb{1}\left[\sigma\left(r^{(c)}_i\right) > 0.5\right]}.
\]
As shown in Table \ref{tab:tfco_results}, in a setting with a sequence length $s$ of 10 seconds, where vehicles only need to appear once within the sequence to be classified as recoverable, 78.3\% of the vehicles are successfully recovered with a mean Euclidean distance of $2.08$ meters from their actual positions. When the architecture focuses on recovering vehicles that appear more frequently within the sequence, the performance improves both in terms of the percentage of recoverable vehicles and the accuracy of their recovered positions. For instance, with a minimum appearance threshold ($a_{\text{min}}$) of 5, the model achieves an 84.5\% recovery rate with a reduced mean distance of $1.66$ meters. Furthermore, the model is also capable of functioning with longer sequences, as demonstrated by the results for a sequence length of $20$ seconds. However, the lower performance in both $a$ and $\overline{e}$ for longer sequences highlights the added difficulty of recovering vehicles under these conditions. This indicates that the ratio of $s$ to $a_{\text{min}}$ is a critical factor in determining the performance of the network. All these results were obtained with a positional loss weight ($\lambda_{\text{position}}$) of $400$. Additional experiments with weights of $300$ and $500$, while keeping the classification loss weight constant, show that the training process can effectively adjust the trade-off between the number of recoverable vehicles and the acceptable positional error. This flexibility enables fine-tuning of the model to prioritize either recovery rate or positional accuracy, depending on the specific application requirements.
\begin{table}[ht]
\centering
\renewcommand{\arraystretch}{1.5} 
\setlength{\tabcolsep}{8pt} 
\begin{tabular}{|c|c|c|c|c|c|}
\hline
$\boldsymbol{s}$ & $\boldsymbol{\alpha_{\text{min}}}$ & $\boldsymbol{\lambda_{\text{position}}}$ & $\boldsymbol{\lambda_{\text{class}}}$ & $a$ [\%] & $\boldsymbol{\overline{e}}$ [m] \\ \hline\hline
10 & 1 & 400 & 1 & 78.3 & 2.08  \\ \hline
10 & 3 & 400 & 1 & 81.0 & 1.85 \\ \hline
10 & 5 & 400 & 1 & 82.5 & 1.66 \\ \hline\hline
5  & 3 & 400 & 1 & 84.6 & 1.3  \\ \hline
20 & 3 & 400 & 1 & 69.9 & 2.32 \\ \hline\hline
10 & 3 & 300 & 1 & 81.9 & 1.89 \\ \hline
10 & 3 & 500 & 1 & 80.1 & 1.82 \\ \hline
\end{tabular}
\caption{Impact of parameters \(s\), \(\alpha_{\text{min}}\), \(\lambda_{\text{position}}\), and \(\lambda_{\text{class}}\) on accuracy (\(a\)) and mean error (\(\overline{e}\)).}
\label{tab:tfco_results}
\end{table}

We compare the results with ablations on the masking applied in the self-attention process. First, we evaluate our proposed masking against no masking, where vehicle embeddings can attend to all other embeddings across the entire sequence. Additionally, we introduce a type encoding to the vehicle embeddings, assigning the same learnable encoding to the same vehicle across different timesteps. Finally, we test a stricter masking approach, allowing embeddings to attend only to the same vehicle across timesteps, but not to other vehicles within the same timestep, effectively reducing the sequence length for each embedding to $n=s$. We evaluated the additional model architectures using the same dataset as before, with parameters \(s=10\) and \(\alpha_{\text{min}}=3\). The results indicate that the model with attention to all other embeddings but without type encoding fails, achieving an accuracy of 0\%. However, incorporating type encodings improves the results to \(a=79.0\%\) and \(\overline{e}=2.42\), demonstrating basic functionality. When attention is restricted to only the same vehicle across the sequence, the model achieves \(a=78.0\%\) and \(\overline{e}=1.96\). All three alternative attention mechanisms yield worse results compared to the proposed method. Attending to vehicles in the same timestep and to the same vehicle embedding across the time sequence strikes a good balance between sparsity and the availability of sufficient information for effective training. Notably, allowing attention to vehicles at the same timestep enables the model to evaluate and account for the continuation of traffic flow, which appears to be crucial for its performance.

\section{Conclusion}
In this work, we explored the concept of FCOs in ITS, highlighting how FCOs can extend FCD by detecting other vehicles through onboard sensors. We introduced multiple methods to model these detection capabilities within microscopic traffic simulations, ranging from light 2D/3D raytracing to high-fidelity co-simulation, and discussed how they each balance accuracy and runtime efficiency. We also presented an emulation technique that uses neural networks to replicate high-fidelity detection results while significantly increasing the section speed and eliminating the need to set up a high-fidelity simulation. Thus, achieving a scalable approximation of realistic detection estimation given a specific sensor setup and 3D detection algorithm. In doing so, this work contributes to existing research by employing detailed modeling of ground truth through high-fidelity simulations, implementing multiple detection techniques for comprehensive comparison with current approaches, and utilizing state-of-the-art machine learning architectures. Our findings show that FCO-based xFCD can substantially enhance the quality of microscopic traffic information compared to traditional FCD alone. Beyond current detections, we demonstrated that FCOs can leverage historical observations to recover vehicles that are no longer directly visible. Incorporating such temporal insights further enriches the traffic state information, enabling the recovery of a significant percentage of previously observed vehicles at the current time step. The proposed transformer-based architecture handles positional information of detected vehicles of the FCOs over multiple time steps and recovers out-of-view vehicles at high accuracy and low position error. Overall, our results underscore the promise of FCO-based approaches for enhanced, fine-grained traffic monitoring and control, offering the potential for future ITS applications.
\newline
\newline
Our primary objective for future research is to extend and refine detection modeling for diverse traffic participants, including VRUs. In this work, we demonstrated our approach using a simplified setup, focusing on vehicle-only detection without significant variation in vehicle classes. However, both microscopic and high-fidelity simulations inherently support multiple road user categories, and the utilized 3D object detectors are capable of handling various classes of traffic participants. Moreover, the emulation approach has already proven effective in more heterogeneous traffic scenarios \cite{gerner2023sumodetector}. Another key direction involves evaluating whether the emulation method can be generalized beyond binary classification to learn bounding boxes directly from BEV representations, which would broaden the potential applications, particularly in automated driving tasks where precise object localization is essential. In this context, we also aim to investigate how emulation models can be adapted or made robust for various weather conditions. Additionally, to better mimic real-world temporal insights, future research will focus on how the recovery process behaves under varying vehicle driving behaviors. In this study, vehicle trajectories were predominantly generated by the SUMO simulator, which may limit diversity in maneuver patterns. Further investigations using richer or mixed driving models will help validate the robustness of our recovery approach. Finally, future research should intend to assess the impact of fine-grained FCO data and temporal insights on a range of ITS applications. One promising example is a Deep Reinforcement Learning (DRL)-based traffic signal control system, which has shown considerable potential for optimizing traffic flow \cite{noaeen2022reinforcement} and is already being piloted in real-world settings \cite{muller2021towards, meess2024first}. 
\section*{Acknowledgement}
This work was partially funded by the Bavarian state government as part of the High Tech Agenda and BayWISS.

\bibliographystyle{elsarticle-num} 
\bibliography{references}

\begin{thebibliography}{10}
\expandafter\ifx\csname url\endcsname\relax
  \def\url#1{\texttt{#1}}\fi
\expandafter\ifx\csname urlprefix\endcsname\relax\def\urlprefix{URL }\fi
\expandafter\ifx\csname href\endcsname\relax
  \def\href#1#2{#2} \def\path#1{#1}\fi

\bibitem{jing2017adaptive}
P.~Jing, H.~Huang, L.~Chen, An adaptive traffic signal control in a connected vehicle environment: A systematic review, Information 8~(3) (2017).
\newblock \href {https://doi.org/10.3390/info8030101} {\path{doi:10.3390/info8030101}}.

\bibitem{huang2018eco}
Y.~Huang, E.~C. Ng, J.~L. Zhou, N.~C. Surawski, E.~F. Chan, G.~Hong, Eco-driving technology for sustainable road transport: A review, Renewable and Sustainable Energy Reviews 93 (2018) 596--609.
\newblock \href {https://doi.org/https://doi.org/10.1016/j.rser.2018.05.030} {\path{doi:https://doi.org/10.1016/j.rser.2018.05.030}}.

\bibitem{schlamp2023glosa}
A.-L. Schlamp, J.~Gerner, K.~Bogenberger, S.~Schmidtner, User-centric green light optimized speed advisory with reinforcement learning, in: 2023 IEEE 26th International Conference on Intelligent Transportation Systems (ITSC), 2023, pp. 3463--3470.
\newblock \href {https://doi.org/10.1109/ITSC57777.2023.10422501} {\path{doi:10.1109/ITSC57777.2023.10422501}}.

\bibitem{tyagi2022routing}
N.~Tyagi, J.~Singh, S.~Singh, A review of routing algorithms for intelligent route planning and path optimization in road navigation, Recent Trends in Product Design and Intelligent Manufacturing Systems: Select Proceedings of IPDIMS 2021 (2022) 851--860.

\bibitem{2023kloekereconomic}
L.~Kloeker, G.~Joeken, L.~Eckstein, Economic analysis of smart roadside infrastructure sensors for connected and automated mobility, in: 2023 IEEE 26th International Conference on Intelligent Transportation Systems (ITSC), 2023, pp. 2331--2336.
\newblock \href {https://doi.org/10.1109/ITSC57777.2023.10422500} {\path{doi:10.1109/ITSC57777.2023.10422500}}.

\bibitem{2023shanlocalization}
X.~Shan, A.~Cabani, H.~Chafouk, A survey of vehicle localization: Performance analysis and challenges, IEEE Access 11 (2023) 107085--107107.
\newblock \href {https://doi.org/10.1109/ACCESS.2023.3318885} {\path{doi:10.1109/ACCESS.2023.3318885}}.

\bibitem{SAEJ3016}
\href{https://www.sae.org/standards/content/j3016\_202104/}{Taxonomy and definitions for terms related to driving automation systems for on-road motor vehicles}~(J3016\_202104) (2021).
\newline\urlprefix\url{https://www.sae.org/standards/content/j3016\_202104/}

\bibitem{elliott2019cav}
D.~Elliott, W.~Keen, L.~Miao, Recent advances in connected and automated vehicles, journal of traffic and transportation engineering (English edition) 6~(2) (2019) 109--131.

\bibitem{ma2021high}
W.~Ma, S.~Qian, High-resolution traffic sensing with probe autonomous vehicles: A data-driven approach, Sensors 21~(2) (2021) 464.

\bibitem{florin2022real}
R.~Florin, S.~Olariu, Real-time traffic density estimation: Putting on-coming traffic to work, IEEE Transactions on Intelligent Transportation Systems 24~(1) (2022) 1374--1383.

\bibitem{zhang2023novel}
Y.~Zhang, M.~Ilic, K.~Bogenberger, A novel concept of traffic data collection and utilization: Autonomous vehicles as a sensor, in: 2023 {{IEEE}} 26th {{International Conference}} on {{Intelligent Transportation Systems}} ({{ITSC}}), 2023.

\bibitem{KITTI}
A.~Geiger, P.~Lenz, R.~Urtasun, Are we ready for autonomous driving? the kitti vision benchmark suite, in: Conference on Computer Vision and Pattern Recognition (CVPR), IEEE, 2012, pp. 3354--3361.

\bibitem{sun2020scalability}
P.~Sun, H.~Kretzschmar, X.~Dotiwalla, A.~Chouard, V.~Patnaik, P.~Tsui, J.~Guo, Y.~Zhou, Y.~Chai, B.~Caine, et~al., Scalability in perception for autonomous driving: Waymo open dataset (2020) 2446--2454.

\bibitem{nuscenes}
H.~Caesar, V.~Bankiti, A.~H. Lang, S.~Vora, V.~E. Liong, Q.~Xu, A.~Krishnan, Y.~Pan, G.~Baldan, O.~Beijbom, \href{http://arxiv.org/abs/1903.11027}{nuscenes: {A} multimodal dataset for autonomous driving}, CoRR abs/1903.11027 (2019).
\newblock \href {http://arxiv.org/abs/1903.11027} {\path{arXiv:1903.11027}}.
\newline\urlprefix\url{http://arxiv.org/abs/1903.11027}

\bibitem{SUMO2018}
P.~A. Lopez, M.~Behrisch, L.~Bieker-Walz, J.~Erdmann, Y.-P. Fl{\"o}tter{\"o}d, R.~Hilbrich, L.~L{\"u}cken, J.~Rummel, P.~Wagner, E.~Wie{\ss}ner, \href{https://elib.dlr.de/124092/}{Microscopic traffic simulation using sumo}, in: The 21st IEEE International Conference on Intelligent Transportation Systems, IEEE, 2018.
\newline\urlprefix\url{https://elib.dlr.de/124092/}

\bibitem{Ilic2024raytracing}
M.~Ilic, M.~Pechinger, T.~Niels, E.~Nexhipi, K.~Bogenberger, An open-source framework for evaluating cooperative perception in urban areas (08 2024).
\newblock \href {https://doi.org/10.13140/RG.2.2.21224.89602} {\path{doi:10.13140/RG.2.2.21224.89602}}.

\bibitem{gerner2023sumodetector}
J.~Gerner, D.~Rößle, D.~Cremers, K.~Bogenberger, T.~Schön, S.~Schmidtner, Enhancing realistic floating car observers in microscopic traffic simulation, in: 2023 IEEE 26th International Conference on Intelligent Transportation Systems (ITSC), 2023, pp. 2396--2403.
\newblock \href {https://doi.org/10.1109/ITSC57777.2023.10422398} {\path{doi:10.1109/ITSC57777.2023.10422398}}.

\bibitem{CARLA2017}
A.~Dosovitskiy, G.~Ros, F.~Codevilla, A.~M. L{\'{o}}pez, V.~Koltun, \href{http://arxiv.org/abs/1711.03938}{{CARLA:} an open urban driving simulator}, CoRR abs/1711.03938 (2017).
\newblock \href {http://arxiv.org/abs/1711.03938} {\path{arXiv:1711.03938}}.
\newline\urlprefix\url{http://arxiv.org/abs/1711.03938}

\bibitem{nozarian2024CARLAkitti}
F.~Nozarian, Carla-kitti, \url{https://github.com/fnozarian/CARLA-KITTI}, accessed: 2024-12-15 (2024).

\bibitem{Liu_Xue_Wu_2022}
X.~Liu, N.~Xue, T.~Wu, \href{https://ojs.aaai.org/index.php/AAAI/article/view/20074}{Learning auxiliary monocular contexts helps monocular 3d object detection}, Proceedings of the AAAI Conference on Artificial Intelligence 36~(2) (2022) 1810--1818.
\newblock \href {https://doi.org/10.1609/aaai.v36i2.20074} {\path{doi:10.1609/aaai.v36i2.20074}}.
\newline\urlprefix\url{https://ojs.aaai.org/index.php/AAAI/article/view/20074}

\bibitem{lang2019pointpillars}
A.~H. Lang, S.~Vora, H.~Caesar, L.~Zhou, J.~Yang, O.~Beijbom, Pointpillars: Fast encoders for object detection from point clouds, in: Proceedings of the IEEE/CVF conference on computer vision and pattern recognition, 2019, pp. 12697--12705.

\bibitem{DBSCAN}
M.~Ester, H.-P. Kriegel, J.~Sander, X.~Xu, et~al., A density-based algorithm for discovering clusters in large spatial databases with noise, in: kdd, Vol.~96, 1996, pp. 226--231.

\bibitem{harth2021automated}
M.~Harth, M.~Langer, K.~Bogenberger, Automated calibration of traffic demand and traffic lights in sumo using real-world observations, in: SUMO Conference Proceedings, Vol.~2, 2021, pp. 133--148.

\bibitem{RESNET}
K.~He, X.~Zhang, S.~Ren, J.~Sun, \href{http://arxiv.org/abs/1512.03385}{Deep residual learning for image recognition}, CoRR abs/1512.03385 (2015).
\newblock \href {http://arxiv.org/abs/1512.03385} {\path{arXiv:1512.03385}}.
\newline\urlprefix\url{http://arxiv.org/abs/1512.03385}

\bibitem{ViT}
A.~Dosovitskiy, L.~Beyer, A.~Kolesnikov, D.~Weissenborn, X.~Zhai, T.~Unterthiner, M.~Dehghani, M.~Minderer, G.~Heigold, S.~Gelly, J.~Uszkoreit, N.~Houlsby, \href{https://arxiv.org/abs/2010.11929}{An image is worth 16x16 words: Transformers for image recognition at scale}, CoRR abs/2010.11929 (2020).
\newblock \href {http://arxiv.org/abs/2010.11929} {\path{arXiv:2010.11929}}.
\newline\urlprefix\url{https://arxiv.org/abs/2010.11929}

\bibitem{IV2024}
J.~Gerner, K.~Bogenberger, S.~Schmidtner, Temporal enhanced floating car observers, in: 2024 IEEE Intelligent Vehicles Symposium (IV), 2024, pp. 1035--1040.
\newblock \href {https://doi.org/10.1109/IV55156.2024.10588538} {\path{doi:10.1109/IV55156.2024.10588538}}.

\bibitem{noaeen2022reinforcement}
M.~Noaeen, A.~Naik, L.~Goodman, J.~Crebo, T.~Abrar, Z.~S.~H. Abad, A.~L. Bazzan, B.~Far, Reinforcement learning in urban network traffic signal control: A systematic literature review, Expert Systems with Applications (2022) 116830.

\bibitem{muller2021towards}
A.~M{\"u}ller, V.~Rangras, T.~Ferfers, F.~Hufen, L.~Schreckenberg, J.~Jasperneite, G.~Schnittker, M.~Waldmann, M.~Friesen, M.~Wiering, Towards real-world deployment of reinforcement learning for traffic signal control, in: 2021 20th IEEE International Conference on Machine Learning and Applications (ICMLA), IEEE, 2021, pp. 507--514.

\bibitem{meess2024first}
H.~Meess, J.~Gerner, D.~Hein, S.~Schmidtner, G.~Elger, K.~Bogenberger, First steps towards real-world traffic signal control optimisation by reinforcement learning, Journal of Simulation (2024) 1--16.

\end{thebibliography}

\end{document}